\documentclass[twocolumn, aps, pre]{revtex4-2}
\usepackage{amsfonts, amsmath, amssymb, pgfplots, float, mathtools, physics }
\usepackage[colorlinks=true,linkcolor=blue,citecolor=blue, urlcolor=blue]{hyperref}
\usepackage{comment}
\usepackage[normalem]{ulem}
\DeclareUnicodeCharacter{2212}{-}
\usepackage{graphicx}
\usepackage{dcolumn}
\usepackage{bm}
\usepackage{ulem}
\usepackage{tabularx}
\usepackage{hyperref}
\usepackage{colortbl,xcolor}

\def\lamt {\tilde{\lambda}}
\newcommand{\cq}{\mathcal{Q}}
\newcommand{\cj}{\mathcal{J}}

\newcommand{\abr}[1]{\left\langle #1 \right\rangle}
\newcommand{\br}[1]{\left( #1 \right)}

\begin{document}

\begin{abstract}

We study steady-state dynamic fluctuations of current and mass, as well as the corresponding power spectra, in a broad class of conserved-mass transport processes on a ring of $L$ sites. These processes violate detailed balance in the bulk and have nontrivial spatial structures: Their steady states are not described by the Boltzmann-Gibbs distribution and, in most cases, are not a-priori known.
By using a microscopic approach, we exactly calculate, for all times $T$, the fluctuations $\langle \mathcal{Q}_i^2(T) \rangle$ and $\langle \mathcal{Q}_{sub}^2(l, T) \rangle$ of the cumulative currents upto time $T$ across $i$th bond and across a subsystem of size $l$ (summed over bonds in the subsystem), respectively. We also calculate the (two-point) dynamic correlation function for subsystem mass.
In particular, we show that, for large $L \gg 1$, the fluctuation $\langle \mathcal{Q}_i^2(T) \rangle$ of the cumulative current up to time $T$ across $i$th bond grows first linearly as $\langle \mathcal{Q}_i^2 \rangle \sim T$ for $T \sim {\cal O}(1)$, subdiffusively as $\langle Q_i^2 \rangle \sim T^{1/2}$ for $1 \ll T \ll L^2$ and then again linearly as $\langle \mathcal{Q}_i^2 \rangle \sim T$ for $T \gg L^2$.
The scaled subsystem current fluctuation $\lim_{l \rightarrow \infty, T \rightarrow \infty} \langle \mathcal{Q}_{sub}^2(l, T) \rangle/2lT$ converges to the density-dependent particle mobility $\chi$ when the large subsystem size limit is taken first, followed by the large time limit; when the limits are reversed, it simply vanishes.
Remarkably, regardless of the models' dynamical rules, the scaled current fluctuation $D \langle \mathcal{Q}_i^2(T)\rangle/2 \chi L \equiv {\cal W}(y)$ as a function of scaled time $y=DT/L^2$ can be expressed in terms of a universal scaling function ${\cal W}(y)$, where $D$ is the bulk-diffusion coefficient; interestingly, the intermediate-time subdiffusive and long-time diffusive growths can be connected through the single scaling function ${\cal W}(y)$. Also, the power spectra for current and mass time series are characterized by the respective scaling functions, which are calculated exactly.
Furthermore, we provide a microscopic derivation of equilibrium-like Green-Kubo and Einstein relations, that connect the steady-state current fluctuations to an ``operational'' mobility (i.e., the response to an external force field)  and mass fluctuation, respectively.

\end{abstract}

\title{Dynamic fluctuations of current and mass in nonequilibrium mass transport processes}

\author{Animesh Hazra}
\author{Anirban Mukherjee}
\author{Punyabrata Pradhan}
\affiliation{Department of Physics of Complex Systems, S. N. Bose National Centre for Basic Sciences, Block-JD, Sector-III, Salt Lake, Kolkata, 700106, India}

\maketitle

\section{Introduction}

\label{Introduction}

Characterizing static and dynamic properties of mass transport processes is a fundamental problem in nonequilibrium statistical physics; it helps develop a simple theoretical understanding of a variety of natural phenomena involving rather complex many-body interactions among constituents, that facilitate transport of mass and energy in a far-from-equilibrium setting. Such processes are abundant in nature and manifest themselves in cloud formation~ \cite{Friedlander_1977}, heat conduction \cite{Kipnis1982Jan}, propagation of forces in granular media~ \cite{Liu_1995,Coppersmith_1996}, river network formation~ \cite{Scheidegger1967Mar}, self-assembly of lipid droplets on cell surfaces~ \cite{Dutta2015}, traffic flow~ \cite{CHOWDHURY2000}, and wealth distribution in a population~ \cite{Yakovenko2009}, among others. A widely studied class of minimal lattice models for understanding transport in interacting-particle systems is that of simple exclusion processes (SEPs) and zero-range processes (ZRPs). Another class of models, which have drawn significant attention in the past, is that of the conserved-mass transport processes, also called {\it mass chipping models} (MCMs)  \cite{Kipnis1982Jan, Liu_1995, Coppersmith_1996, Rajesh2000_jstat, Garcia2000, Zielen2002, Bondyopadhyay_2012, das_spatial_2016}. Interestingly, their steady-state measures on a closed geometry, unlike that for SEPs and ZRPs, are not usually described by the equilibrium Boltzmann-Gibbs distribution and, in most cases, are {\it a-priori} not known. Indeed, these systems are inherently driven far from equilibrium and generate nontrivial spatial structures, making exact dynamic characterization of steady-state fluctuations a challenging problem.

Recently, a theoretical framework for driven diffusive systems, known as macroscopic fluctuation theory (MFT) \cite{Derrida2007Jul, Bertini2015Jun}, has been developed for studying fluctuations of coarse-grained (hydrodynamic) variables such as density $\rho(x,\tau)$ and current $j(x,\tau)$, where $x$ and $\tau$ are suitably rescaled position and time, respectively.
The MFT is a generalization of the Onsager-Machlup theory of {\it near-equilibrium systems} to the theory of far-from-equilibrium ones \cite{Onsager1953Sep, Landau}.
Its main ingredients are the density-dependent transport coefficients, namely the bulk-diffusion coefficient $D(\rho)$ and the mobility $\chi(\rho)$ (equivalently, the conductivity), which govern density relaxation and current fluctuation on macroscopic scales \cite{Bertini2001Jul, Bodineau2004May, Hurtado2009Jun, Rizkallah2023Jan}.
Despite a simple prescription of the MFT, calculating the transport coefficients as a function of density, and other parameters, is difficult, especially for many-body systems, where spatial correlations are nonzero.
The difficulty stems primarily from the fact that the averages of various observables, which are necessary to calculate the transport coefficients, must be computed in the nonequilibrium steady state, which is not described by the Boltzmann-Gibbs distribution and, moreover, is not explicitly known in most cases.
Perhaps not surprisingly, apart from SEPs \cite{Derrida1998Jan, Derrida2004May, Derrida2009Jul, Gabrielli2018Apr, Fiore2021Dec, Grabsch2023Apr, Grabsch2022Mar} and ZRPs \cite{Harris2005Aug, Bertini2006Apr}, which have a product-measure steady state \cite{Blythe2007Oct}, there are very few examples of exact microscopic characterization of dynamic fluctuations in interacting-particle systems.

Of course, MCMs, which constitute a paradigm for out-of-equilibrium many-body systems, are an exception. 
Indeed, because they are analytically tractable, MCMs provide a level playing field for exact microscopic calculations of various time-dependent quantities, such as static density correlations and dynamic tagged-particle correlations, which have been extensively explored in the past \cite{Rajesh2000_jstat, das_spatial_2016, Cividini2016Jan}.
However, except for the Kipnis-Marchioro-Presutti (KMP)-like models \cite{Prados2012Sep, Basile2006May} and the SEP \cite{Sadhu_2016}, which satisfy detailed balance, rigorous microscopic  characterization of the precise quantitative connection between fluctuation and transport have not been done for nonequilibrium mass-transport processes having nontrivial spatial structures in the bulk.
Indeed it would be quite interesting to relate the microscopic dynamic properties of mass and current to the macroscopic transport coefficients and thus to derive the MFT \cite{Bertini2015Jun} for such models from ``first-principles'' calculations.

In this paper, by using a microscopic approach, we exactly calculate dynamic correlations for subsystem current and mass in a broad class of one-dimensional mass chipping models (MCMs) on a ring of $L$ sites.
In these models, a site $i$ contains a continuous mass $m_i \ge 0$ and the total mass in the systems remains conserved. With some specified rates, a certain fraction of mass at a site gets fragmented or chipped off from the parent mass, diffuse {\it symmetrically} and coalesce with mass at one of the nearest-neighbor sites. The MCMs have been intensively studied in various contexts in the past decades \cite{Rajesh2000_jstat, Garcia2000, Bondyopadhyay_2012, das_spatial_2016,  Cividini2016Jan,  Dandekar2023Jan}, and can be mapped to a class of transport processes, called the \textit{random averaging processes} (RAPs) \cite{Ferrari1998}, which is again a variant of the so called Hammersley process \cite{Aldous1995Jun}. Note that, for symmetric transfer (i.e., diffusion) of masses, although there is no net mass flow in the steady state on a ring geometry, the probability currents in the configuration space can still be nonzero and the Kolmogorov criteria for equilibrium can be shown to get violated \cite{Arghya_2017}. 
As mentioned before, despite the steady-state measures for generic parameter values are not known  \cite{Garcia2000, Rajesh2000_jstat, Bondyopadhyay_2012, Cividini2016Jan}, the MCMs are amenable to exact theoretical studies. For example, the spatial correlation function of mass has been exactly calculated before in some of the variants of MCMs \cite{Rajesh2000_jstat, Rajesh_2000_PRE, Rajesh_2001_PRE, Garcia2000, Zielen2002}.
Furthermore,  the mean-squared fluctuation of the position of a single tagged particle as well as the dynamic correlations of two tagged particles in related models - the RAPs - have been calculated exactly using microscopic and hydrodynamic calculations  \cite{Rajesh_2001_PRE, Schutz2000May, Kundu2016Oct}.

The primary focus of our study is the cumulative time-integrated currents -  $\mathcal{Q}_i(T)$ and $\mathcal{Q}_{sub}(l, T)$ in a time interval $[0,T]$ across a bond $(i, i+1)$ and a subsystem of size $l$, respectively. The bond current fluctuation $\langle {\cal Q}^2_i(T) \rangle$ as a function of time $T$ exhibits three distinct temporal behaviors. Initially, for small times $T \ll 1/D$, the temporal growth is linear in time $T$, where $D$ is the bulk-diffusion coefficient (a constant). At moderately large times, the fluctuation grows subdiffusively, having a $T^{1/2}$ growth for $1/D \ll T \ll L^2 / D$ with $L$ being the system size. Finally, at very large times $T \gg L^2/D$, the growth again becomes linear in time. We find that, even in the presence of nonzero spatial correlations, the qualitative behaviour of the current fluctuations, except for the prefactors, have characteristics, which are similar to that in the SEP.
Remarkably, independent of the details of the mass transfer rules of the models,  the suitably scaled bond-current fluctuation $\langle \mathcal{Q}_i^2(T) \rangle D / 2\chi L$, with $\chi(\rho)$ is the density-dependent mobility, as a function of the scaled time $y=DT/L^2$ can be expressed in terms of a universal scaling function $\mathcal{W}(y)$, which is exactly calculated and is shown to have the following asymptotic behavior,
\begin{align}
    \label{eq:scaling_function_form}
    \mathcal{W}(y) = 
    \begin{cases}
        \left( \frac{y}{\pi} \right)^{1/2} & {\rm for} \hspace{4pt} y \ll 1, \\
        y & {\rm for} \hspace{4pt} y \gg 1.
    \end{cases}
\end{align}
This demonstrates that the intermediate-time $T^{1/2}$ and long-time $T$ growths of bond-current fluctuation can be related through a single scaling function ${\cal W}(y)$.
Furthermore, we show that the two-point correlation for the instantaneous current as a function of time $t$ has a delta correlated part at $t=0$ and a long-ranged (power law) negative part, which decays as $t^{-3/2}$. The corresponding power spectrum of current $S_{\mathcal{J}}(f)$ is calculated analytically and it exhibits a low-frequency power-law behavior $f^{1/2}$ in the frequency regime $D/L^2 \ll f \ll 1$. Similarly, the power spectrum $S_{M_l}(f)$ for the subsystem mass is calculated exactly and is shown to have a low-frequency power-law divergence $f^{-3/2}$. We have also calculated the scaling functions when the rescaled power spectra for current and mass are expressed in terms of the scaled frequency $fL^2/D$.

We derive a nonequilibrium fluctuation relation between scaled subsystem mass and space-time integrated current fluctuations. We calculate the scaled fluctuation of the cumulative current $\mathcal{Q}_{sub}(l,T)$,  summed over a subsystem of size $l$ and integrated up to time $T$, and we show that the scaled subsystem current fluctuation converges to the density-dependent particle mobility $\chi(\rho)$, i.e., a nonequilibrium Green-Kubo-like formula,
\begin{equation}
\sigma^2_{\mathcal{Q}}\equiv\lim_{l \rightarrow \infty, T \rightarrow \infty} \frac{ \langle\mathcal{Q}_{sub}^2(l,T) \rangle}{ lT} = 2\chi(\rho),
\end{equation}
where the infinite subsystem-size limit $l \rightarrow \infty$ is taken first, followed by the infinite time limit $T \rightarrow \infty$; notably, in the opposite order of limits, the lhs simply vanishes.
By explicitly calculating the scaled subsystem mass fluctuation $\sigma^2_{M}=\lim_{l\to\infty}\langle\Delta M_l^2\rangle /l$, where $\langle\Delta M_l^2\rangle = \langle M_l^2 \rangle-\langle M_l \rangle^2$ is the fluctuation of mass in a subsystem of size $l$, we then derive a nonequilibrium fluctuation relation between mass and current fluctuations,
\begin{align}
    \sigma^2_{M}=\frac{\sigma^2_{\mathcal{Q}}}{2D},
\end{align}
which is a modified version of the celebrated Einstein relation for equilibrium systems. Furthermore, provided there is a small biasing force $\tilde F$ (suitably scaled), which generates a drift current $J_{drift} = \chi_{op}(\rho) \tilde{F} $ along the direction of the force, we derive a Green-Kubo-like fluctuation-response relation,
\begin{equation}
 \chi_{op}(\rho) \equiv \left[ \frac{\partial J_{drift}}{\partial \tilde F} \right]_{\tilde F=0} = \frac{\sigma^2_Q}{2};
\end{equation}
the above relation directly connects the ``operational mobility'' or, equivalently, the response, due to the applied force, to the current fluctuations in the nonequilibrium steady state.

We organize the rest of the paper as follows. In section \ref{Sec:model} we introduce three mass chipping models: MCM I, MCM II, and MCM III. In Section \ref{Sec:Theory}, we calculate the dynamic correlations, time-integrated bond current fluctuations, and power spectra of instantaneous bond current and subsystem mass fluctuations within the context of MCM I. In sections \ref{Sec:Model_II} and \ref{Sec:MCM_III} we present the calculations for dynamic correlations for the other two models, MCM II and MCM III. We perform a thorough comparative analysis of the dynamic properties exhibited by these models in section \ref{Sec:Comparison}. Finally, in Sec. \ref{Sec:Conclusion}, we summarize our results with some concluding remarks.

\section{Models}
\label{Sec:model}

\begin{widetext}
\begin{figure*}
    \centering
    \includegraphics[scale=0.45]{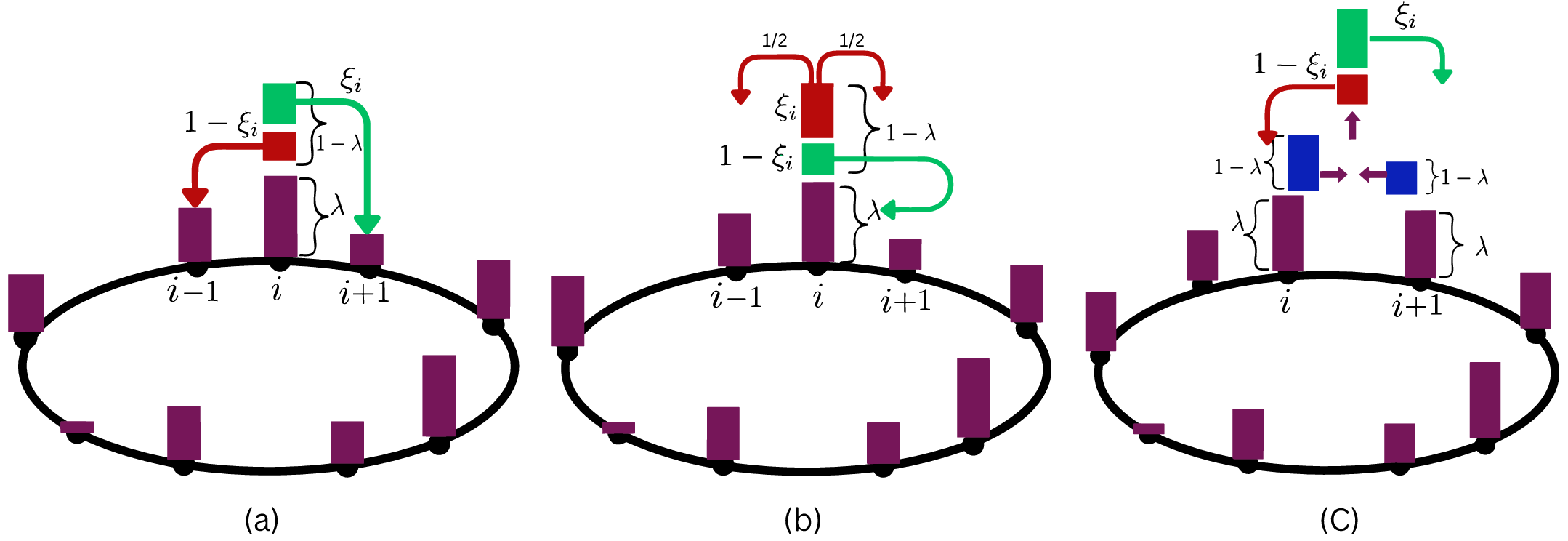}
    \caption{\textit{Schematic representation of the mass-chipping models MCM I, MCM II, and MCM III:} (a) In MCM I, a site $i$ on a periodic lattice (shaped as a dark oval) having mass $m_i$ (dark violet), retains a fraction $\lambda$ of its mass (dark violet), while a random fraction $\xi_i$ of the chipped-off mass (green) migrates to the right neighbor, and the remaining fraction of the chipped mass(red) moves to the left side. (b) In MCM II, a random fraction $\xi_i$ of chipped-off mass(red) moves either left or nearest neighbor with equal probability, while the rest of the chipped mass(green) is deposited back to the same site $i$. (c) In MCM III, a fraction equivalent to $1-\lambda$ of the mass (blue) is chipped off from sites $i$ and $i+1$. This extracted mass is then recombined and subsequently redistributed in a way such that site $i$ receives a random fraction of $1-\xi_i$ (red), while site $i+1$ acquires a fraction of $\xi_i$ (green).}
    \label{fig:model_schematic}
\end{figure*}
\end{widetext}

In this section, we define three well studied variants of conserved mass chipping models: MCM I, MCM II, and MCM III, which differ in the details of their microscopic dynamics. We define the models in a one-dimensional periodic lattice with sites labeled by $i=0,1,\cdots, L-1$. A mass $m_i\geq 0$ is associated with each site $i$, with the total mass $M=\sum_{i=0}^L m_i$ being conserved. In these models, we introduce a variable $\lambda$ that determines the fraction of mass retained by the site and other fraction of mass $\tilde{\lambda}=1-\lambda$, called chipping constant, which gets chipped off from the parent mass. 
In Fig. \ref{fig:model_schematic} we present schematic diagrams that represent the underlying microscopic dynamics of these three models. The continuous-time dynamical update rules are provided below. 
\\
\\
\textbf{MCM I: }
 A site is updated with unit rate. A constant $\lambda$ fraction of mass is retained at the site, while $\lamt = 1-\lambda$ fraction of mass $m_i$ is chipped off. Subsequently, a random fraction $\xi_i$ of the chipped-off mass, i.e., $\lamt m_i \xi_i$, is then transferred to the right nearest neighbor, while the remaining fraction of the chipped-off mass, $\lamt m_i (1-\xi_i)$, is transferred to the left nearest neighbor. Here $\xi_i\in [0,1]$ is independent and identically distributed (i.i.d.) random variables, having a uniform distribution. For convenience, we also define $\tilde{\xi}_i=1-\xi_i$ for later use.
\\
\\ 
\textbf{MCM II:}
A specific fraction, $\lambda$, is retained, while the fraction, $\lamt=1-\lambda$, is chipped off. Additionally, a random fraction $\xi_i$ of the chipped-off mass, i.e., $\lamt m_i \xi_i$, is then transferred either to the left or to the right, with an equal probability $1/2$. The remaining fraction of the chipped-off mass, $\lamt m_i(1-\xi_i)$, is subsequently deposited back to site $i$.
\\
\\
\textbf{MCM III:} 
In this model, a bond $(i,i+1)$ is updated with unit rate. A fraction $\tilde \lambda=1-\lambda$ of their  masses is chipped off,i.e., resulting in $\lamt m_i$ being removed from the site $i$ and $\lamt m_{i+1}$ from the site $i+1$. These chipped-off masses are then combined and, subsequently, a random fraction, $\xi_i$, of the combined mass is transferred to site $i+1$, while the $\tilde \xi_i = 1-\xi_i$ fraction is transferred to site $i$.

\section{Theory: MCM I}

\label{Sec:Theory}

In this section, we study in detail the first variant of mass-chipping models, i.e., MCM I, on a periodic one-dimensional lattice of size $L$; For other models MCMs II and III, we later state the results, which can be derived following the techniques developed in this section.

A site $i$, with $i=0$, $1$, $\ldots$, $L-1$, possess a  mass $m_i$, which can take continuous value in the interval $0 \leq m_i \leq M$; total mass $M = \sum_{i=0}^{L-1} m_i = M$ remains conserved throughout and is the only conserved quantity. Density is defined as $\rho= M/L$, however we denote $\rho_i(t) = \langle m_i(t) \rangle$ as local density at site $i$ and at time $t$.
Notably, unlike in MCMs II and III, a site in MCM I is stochastically updated in a way that simultaneously impacts its immediate neighbours, as stated in the previous section. This results in nonzero spatial correlations, making the calculations nontrivial.

We can now explicitly write down the stochastic update rules for mass $m_i(t)$ at site $i$ and at time $t$ during an infinitesimal time interval $(t, t+dt)$,
\begin{align}\label{eq:mass_update}
m_i(t+dt)=
    \begin{cases}
        \textbf{event}         & \textbf{prob.} \\
        m_i(t)-\lamt m_i(t)       & dt\\
        m_i(t)+\lamt \xi_{i-1} m_{i-1}(t)   & dt\\
        m_i(t) + \lamt \tilde{\xi}_{i+1} m_{i+1}(t)   & dt\\
        m_i(t)  & (1-3dt),
        \end{cases}
\end{align}
where $\xi_j \in (0,1)$ is a random variable, which, for simplicity, is taken to be uniformly distributed; generalization of the results to other distributions is straightforward. Using the above dynamical update rules, the time-evolution of  local mass can be written as
\begin{align}
    \label{eq:generic_diff_eqn}
    \frac{d}{dt} \abr{m_i(t)} = D(\lambda) \br{ \abr{m_{i-1}(t)} - 2\abr{m_{i}(t)} + \abr{m_{i+1}(t)} },
\end{align}
where $D(\lambda)=\tilde{\lambda}/2$ is the bulk-diffusion coefficient for MCM I. Note that $D$ is independent of density, leading to some important simplifications in the hierarchy for mass and current correlation functions, which, as we show later, actually close.

\subsection{Definitions and notations}

At this point, we introduce time-integrated bond current $\mathcal{Q}_i(t)$, which is the cumulative current across bond $(i,i+1)$ in the time interval time $(0, t)$. The time-integrated current across the $i^{th}$ bond during an infinitesimal time interval $[t,t+dt]$ is simply $\mathcal{J}_i(t)dt$, where instantaneous bond current
\begin{equation}
\mathcal{J}_i(t) \equiv \frac{d \mathcal{Q}_i(t)}{dt},
\end{equation}
and therefore we have the time-integrated current across bond $(i,i+1)$
\begin{equation}
  \mathcal{Q}_i(t) = \int\limits_{0}^{t}dt^\prime \mathcal{J}_i(t^\prime).
\end{equation}
We can then express Eq.(\ref{eq:generic_diff_eqn}), the time evolution 
of the local density $\rho_i(t)=\langle m_i(t)\rangle$, in terms of a continuity equation for local density $\rho_i(t) = \langle m_i(t) \rangle$ and the average local current $\langle \mathcal{J}_i(t) \rangle$ simply as 
\begin{equation}\label{eq:density}
\frac{d}{dt} \rho_i(t) = \langle \mathcal{J}_i(t) - \mathcal{J}_{i+1}(t) \rangle .
\end{equation}
It is useful to decompose the instantaneous bond current as the sum of a diffusive component $\mathcal{J}_{i}^{(d)}(t)$ and a fluctuating component $\mathcal{J}_{i}^{(fl)}$ as 
\begin{equation}\label{eq:J_it}
    \mathcal{J}_i(t) = \mathcal{J}_{i}^{(d)}(t)+\mathcal{J}_{i}^{(fl)}(t),
\end{equation}
where we can identify the diffusive current $\mathcal{J}_i^{(d)}(t)$ as
\begin{equation}\label{eq:dif_cur}
\mathcal{J}_i^{(d)}(t) \equiv D(\lambda) \big[m_i(t)-m_{i+1}(t)\big].
\end{equation}
The diffusion constant $D(\lambda)$ depends only on the chipping constant $\lamt$, not density $\rho$. It should be noted that the average fluctuating current $\langle \mathcal{J}_i^{(fl)}\rangle = 0$, implying $\langle \mathcal{J}_i(t)\rangle=\langle \mathcal{J}_i^{(d)}(t)\rangle$. Indeed, one could interpret $\mathcal{J}_i^{(fl)}(t) $ as a fast varying ``noise'' current around the slowly varying diffusive (``hydrodynamic'') current component $\mathcal{J}_i^{(d)}(t)$. This decomposition of current is important because, as we show later explicitly, the fluctuation statistics of $\mathcal{J}_i^{(fl)}(t)$ is in fact strictly delta-correlated in time and short-ranged in space, whereas the diffusive current  $\mathcal{J}_i^{(d)}(t)$ is long-ranged in time (in fact, a power law) and short-ranged in space.

For convenience, we introduce the following notation for correlation function $C_r^{AB}(t,t^\prime)$ involving  any two local observable $A_i(t)$ and $B_j(t^\prime)$, with $t\geq t^\prime$,
\begin{equation}
\begin{aligned}
C_{r=|j-i|}^{AB}(t,t^\prime) &= \langle A_i(t)B_j(t^\prime) \rangle - \langle A_i(t)\rangle \langle B_j(t^\prime) \rangle\\
& \equiv \langle A_i(t)B_j(t^\prime) \rangle_c,
\end{aligned}
\end{equation}
where $r=|j-i|$ is the relative distance. We denote the spatial Fourier transform of the correlation function $C_r^{AB}(t,t^\prime)$ as given below
\begin{equation}\label{eq:fourier_def}
\tilde{C}_q^{AB}(t,t^\prime)= \sum\limits_{r=0}^{L-1} C_r^{AB}(t,t^\prime) e^{i qr} ,
\end{equation}
where $q=2\pi s/L$ and $s=0,1, \dots , L-1$; the inverse Fourier transform  is given by
\begin{equation}
C_r^{AB}(t,t^\prime)= \frac{1}{L}\sum\limits_{q} \tilde{C}_q^{AB}(t,t^\prime) e^{-i qr} .
\end{equation}

\subsection{Calculation scheme}

In this section we describe our calculation scheme in details for MCM I.
The stochastic dynamical rules for  time-integrated current $\mathcal{Q}_i(t)$ in an infinitesimal time interval $(t,t+dt)$ can be written as
\begin{align}\label{eq:cur_update}
\mathcal{Q}_i(t+dt)=
    \begin{cases}
        \textbf{event}         & \textbf{prob.} \\
        \mathcal{Q}_i(t)+\lamt \xi_im_i(t)       & dt\\
        \mathcal{Q}_i(t) - \lamt \xi_{i+1} m_{i+1}(t)   & dt\\
        \mathcal{Q}_i(t) & (1-2dt).
        \end{cases}
\end{align}
The above update rules allow us to derive the time-evolution equation for the first moment of the time-integrated bond-current $\mathcal{Q}_i(t)$ as follows:
\begin{align}\label{eq:Q_it}
  \frac{d \langle \mathcal{Q}_i(t)\rangle}{dt}= D\langle m_i(t) - m_{i+1}(t) \rangle = \langle \mathcal{J}_i^{(d)}(t) \rangle.
\end{align}
Using the update rule as in Eq.(\ref{eq:cur_update}), the infinitesimal time-evolution equation for following product of the time-integrated currents at two different times $t$ and $t^\prime$($t > t^\prime $) can be written as
$$\mathcal{Q}_i(t+dt)\mathcal{Q}_{i+r}(t^\prime)=~~~~~~~~~~~~~~~~~~~~~~~~~~~~~~$$
\begin{align}\label{eq:cur_cur_un_update}
    \begin{cases}
        \textbf{event}         & \textbf{prob.} \\
        [\mathcal{Q}_i(t)+\lamt \xi_im_i(t)]\mathcal{Q}_{i+r}(t^\prime)       & dt\\
        [\mathcal{Q}_i(t) - \lamt \xi_{i+1} m_{i+1}(t)]\mathcal{Q}_{i+r}(t^\prime)   & dt\\
        \mathcal{Q}_i(t)\mathcal{Q}_{i+r}(t^\prime) & (1-2dt).
        \end{cases}
\end{align}
Now expressing  Eq.(\ref{eq:j_dm}) in terms of mass and after some algebraic manipulation, we immediately get the following equality,
\begin{equation}
  \label{eq:QQ_ittp_1}
\frac{d}{dt}C_r^{\mathcal{Q}\mathcal{Q}}(t,t^\prime) =C_r^{\cj^{(d)}\mathcal{Q}}(t,t^\prime).
\end{equation}
Interestingly, while calculating time derivative of current (or related observales), Eq. (\ref{eq:QQ_ittp_1})  can be simply obtained by using a convenient thumb rule where one takes the time derivative inside angular brackets as 
\begin{align}
\begin{aligned}\label{eq:j_dm}
  \frac{d}{dt}\left\langle \cq_i(t)\cq_{i+r}(t^\prime)\right\rangle_c & \equiv \left\langle\frac{d\cq_i(t)}{dt}\cq_{i+r}(t^\prime)\right\rangle
  \\
    &-\abr{\frac{d\cq_i(t)}{dt}} \left\langle \cq_{i+r}(t^\prime)\right\rangle.
    \end{aligned} 
\end{align}
Then by replacing the instantaneous current through the equivalence relation ${d\cq_i(t)}/{dt} \equiv D(m_i-m_{i+1}) + \mathcal{J}^{(fl)}$ and subsequently dropping the noise correlation as $\langle {\cal J}_i^{(fl)}(t) {\cal Q}_{i+r}(t') \rangle = 0$ for $t>t'$, we get Eq. (\ref{eq:QQ_ittp_1}).

Now, using Eq. (\ref{eq:dif_cur}) into rhs of Eq.(\ref{eq:QQ_ittp_1}), we can immediately express the time evolution of unequal-space-time current-current correlation function in terms of the unequal-space-time mass-current correlation function,
\begin{equation}\label{eq:qq_t}
\frac{d}{dt}C_r^{\mathcal{Q}\mathcal{Q}}(t,t^\prime) = D\big(C_r^{m\mathcal{Q}}(t,t^\prime)- C_{r-1}^{m\mathcal{Q}}(t,t^\prime)\big).
\end{equation}
From the above equation, we see that we now require to calculate the unequal-time mass-current correlation $C_r^{m\mathcal{Q}}(t,t^\prime)$ in order to determine the unequal-time current-current correlation $C_r^{\mathcal{Q}\mathcal{Q}}(t,t^\prime)$. The time evolution of the correlation function $C_r^{m\mathcal{Q}}(t,t^\prime)$ can be obtained by using infinitesimal-time update rules for the following mass-current product at a later time $t+dt$ as
\\
$$m_i(t+dt)Q_{i+r}(t^\prime)= ~~~~~~~~~~~~~~~~~~~~~~~~~~~~~~~~~~~~~~~~~~~~$$
\begin{align}\label{eq:mass_cur_un_update}
    \begin{cases}
        \textbf{event}         & \textbf{prob.} \\
        [m_i(t)-\lamt m_i(t)]Q_{i+r}(t^\prime)       & dt\\
        [m_i(t)+\lamt \xi_{i-1} m_{i-1}(t)]Q_{i+r}(t^\prime)   & dt\\
        [m_i(t) + \lamt \tilde{\xi}_{i+1} m_{i+1}(t)]Q_{i+r}(t^\prime)   & dt\\
        m_i(t)Q_{i+r}(t^\prime)  & (1-3dt).
        \end{cases}
\end{align}
Using above update rule the time evaluation of unequal-time mass-current can be expressed in the following form,
\begin{align}\label{eq:mq_t}
\begin{aligned}
  &\frac{d}{dt} C_r^{m\cq}(t,t^\prime)
  \\
  &=D\langle (m_{i+1}(t)-2m_i(t)+m_{i-1}(t))\mathcal{Q}_{i+r}(t^\prime)\rangle_c
  \\
&= D \sum\limits_k \Delta_{r,k}C_{k}^{m\cq}(t,t^\prime),
\end{aligned}
\end{align}
where $\Delta_{r,k}=\delta_{r-1,k}-2\delta_{r,k}+\delta_{r+1,k}$ is the discrete Laplacian operator.
Now, equations (\ref{eq:qq_t}) and (\ref{eq:mq_t}) can be expressed in terms of the Fourier modes as defined in Eq.(\ref{eq:fourier_def}),
\begin{equation}\label{qq_t2}
\frac{d}{dt}\tilde{C}_q^{\cq\cq}(t,t^\prime) = D\tilde{C}_q^{m\cq}(t,t^\prime)\left(1-e^{iq}\right),
\end{equation}
and 
\begin{equation}\label{mq_t2}
\frac{d}{dt}\tilde{C}_q^{m\cq}(t,t^\prime)=-D\omega_q\tilde{C}_q^{m\cq}(t,t^\prime),
\end{equation}
where the eigenvalue of discrete Laplacian is written as follows
\begin{equation}\label{eq:omega_q}
\omega_q=2\left(1-\cos q\right).
\end{equation}
Also, $\tilde{C}_q^{\cq\cq}(t,t^\prime)$ and $\tilde{C}_q^{m\cq}(t,t^\prime)$ are the Fourier transforms of the quantities $C_r^{\cq\cq}(t,t^\prime)$ and $C_r^{m\cq}(t,t^\prime)$, respectively.
Now, Eq.(\ref{qq_t2}) and Eq.(\ref{mq_t2}), can be integrated to have
\begin{equation}\label{eq:qq_qt}
\begin{aligned}
\tilde{C}_q^{\cq\cq}(t,t^\prime) &= D\int\limits_{t^\prime}^{t}dt^{\prime\prime}\tilde{C}_q^{m\cq}(t^{\prime\prime},t^\prime)\left(1-e^{iq}\right)\\&+\tilde{C}_q^{\cq\cq}(t^\prime,t^\prime),
\end{aligned}
\end{equation}
and
\begin{equation}\label{eq:mq_qt}
\tilde{C}_q^{m\cq}(t,t^\prime)=e^{-D\omega_q(t-t^\prime )}\tilde{C}_q^{m\cq}(t^\prime,t^\prime) 
\end{equation}
respectively. The equations Eq.(\ref{eq:qq_qt}) and Eq.(\ref{eq:mq_qt}) suggest that the dynamic correlation of current-current $C_r^{m\cq\cq}(t^\prime,t^\prime)$ and mass-current $C_r^{m\cq}(t^\prime,t^\prime)$ at equal time are required to obtain the respective dynamic correlation at unequal time, from their corresponding update rules.

The time-evolution equation for the equal-time current-current spatial correlation $C_r^{\cq\cq}(t,t)$ can be written from the infinitesimal  update rules for the product of the following random variables,
$$\mathcal{Q}_i(t+dt)\mathcal{Q}_{i+r}(t+dt) = ~~~~~~~~~~~~~~~~~~~~~~~~~~~~~~~~~~~~~~~~$$
\begin{align}\label{eq:cur_cur_update}   
   \begin{cases}
     \textbf{event}         & \textbf{prob.} \\
     \mathcal{Q}_i\mathcal{Q}_{i+r}+\lamt\left( \xi_im_i-\tilde{\xi}_{i+1}m_{i+1}\right)\mathcal{Q}_{i+r}\\
     +\lamt\left( \xi_{i+r}m_{i+r}-\tilde{\xi}_{i+r+1}m_{i+r+1}\right)\mathcal{Q}_i & dt \\
     \mathcal{Q}_i\mathcal{Q}_{i+r}+\lamt^2(\xi_i^2m_i^2+\tilde{\xi}_{i+1}^2m_{i+1}^2) & \delta_{r,0}dt \\
     \mathcal{Q}_i\mathcal{Q}_{i+r}-\lamt^2\xi_i\tilde{\xi}_im_i^2 & \delta_{r,-1}dt \\
     \mathcal{Q}_i\mathcal{Q}_{i+r} -\lamt^2\xi_{i+1}\tilde{\xi}_{i+1}m_{i+1}^2 & \delta_{r,1}dt \\
     \mathcal{Q}_i\mathcal{Q}_{i+r} & 1-\sum dt,
   \end{cases}
\end{align}
where $\sum=1+\delta_{r,0}+\delta_{r,1}+\delta_{r,-1}$ represents the total exit rate. Hence, from the above update rules, we can deduce the following time-evolution equation,
\begin{equation}\label{eq:dt_cur}
\begin{aligned}
    \frac{d}{dt}\abr{\cq_i(t)\cq_{i+r}(t)}_c &= D\abr{(m_i-m_{i+1})\cq_{i+r}}_c\\& + D\abr{\cq_i(m_{i+r}-m_{i+r+1})}_c +\Gamma_r,
    \end{aligned}
\end{equation}
where $\Gamma_r$ can be written in terms of steady-state single-site mass fluctuation (function of $\rho$),
\begin{align}\label{eq:gamma_r}
\Gamma_r (\rho) = \frac{\lamt^2}{6}\langle m_i^2\rangle(4\delta_{r,0}-\delta_{r,1}-\delta_{r,-1}) . 
\end{align}
For  convenience, we introduce the following  quantity, 
\begin{align}\label{eq:mobility:mcm_1}
    \chi(\rho) \equiv \frac{\lamt^2}{6}\langle m_i^2\rangle
\end{align}
which, as we show later, is nothing but the density-dependent transport coefficient, called mobility $\chi(\rho)=\lim\limits_{T\to \infty, L\to\infty} L\abr{\cq_i^2(T)}/2T$ - the scaled bond-current fluctuation, with infinite time limit $T\to\infty$ taken first \cite{Derrida2004May}. As we shall demonstrate later, the mobility $\chi(\rho)$ can be exactly equated to another related transport coefficient, we call it ``operational'' mobility $\chi_{op}(\rho)$, which is ratio of the current (response) to a small externally applied biasing force (perturbation) \cite{Arghya_2017}.
The expression for the second moment of mass $\langle m_i^2\rangle$ in the steady state can be written in terms of chipping constant $\lamt$ and the density $\rho$ as given below \cite{das_spatial_2016},
\begin{align}\label{eq:mi2}
\langle m_i^2\rangle = \frac{3\rho^2}{3-2\lamt}.
\end{align}
We can now substitute Eq.(\ref{eq:mobility:mcm_1}) into Eq.(\ref{eq:gamma_r}) to express $\Gamma_r$ in terms of the system's mobility $\chi$ as 
\begin{equation}\label{eq:gamma_r_f}
    \Gamma_r(\rho) = 4\chi(\rho) \delta_{r,0}-\chi (\delta_{r,1}+\delta_{r,-1}).
\end{equation}
It is interesting to note that $\Gamma_r$ has a direct connection to the steady-state mass-mass correlation $C_r^{mm}$, through the relation
\begin{align}\label{eq:gamm_r_cr}
    \Gamma_r = 2DC_r^{mm},
\end{align}
which should be generic for diffusive systems \cite{Anirban_2023, Chakraborty2023Sep}. Later we show that the quantity $\Gamma_r$ is also related to the spatial correlation function for the fluctuating (``noise'') current, thus establishing a direct (presumably generic) connection between (noise) current fluctuation and density fluctuation in a diffusive system and, thus, characterizing the role of steady-state spatial structure in determining the large-scale dynamic properties. This is precisely how density and current fluctuations, as well as relaxation properties (through bulk-diffusivity), are intricately coupled to one another, resulting in an equilibrium-like Einstein relation, as demonstrated subsequently.

Now, by using the following formula
\begin{align}
    \begin{aligned}
        D\abr{(m_i-m_{i+1})\cq_r}_c = \frac{D}{L}\sum\limits_{q}(1-e^{iq})\tilde{C}_q^{m\cq}(t,t)e^{-iqr}
    \end{aligned}
\end{align}
in Eq.(\ref{eq:dt_cur}) and performing some algebraic manipulations, we obtain the following expression,
\begin{align}\label{eq:crt_crq}
    \frac{d}{dt}C_r^{\cq\cq}(t,t) = \frac{D}{L}\sum\limits_{q}(1-e^{iq})\tilde{C}_q^{m\cq}(t,t)(2-\omega_{qr})+ \Gamma_r,
\end{align}
where $\omega_{qr}=2\left(1-\cos(qr)\right)$, as indicated in Eq.(\ref{eq:omega_q}). After integrating both sides of the above equation, we obtain the equal-time current-current dynamic correlation as follows:
\begin{equation}\label{eq:crQQtt}
\begin{aligned}
&C_r^{\cq\cq}(t,t) = \int\limits_{0}^{t}dt^\prime \Gamma_r(t^\prime) +\\ &\frac{D}{L}\int\limits_{0}^{t}dt^\prime \sum\limits_{q}\tilde{C}_q^{m\cq}(t^\prime,t^\prime)(1-e^{iq})(2-\omega_{qr}).
\end{aligned}
\end{equation}
Now, to obtain the desired form of the above equal-time correlation for current, we calculate the equal-time correlation function $C_r^{m\cq}(t,t)$ by using the following infinitesimal-time update rule,
$$m_i(t+dt)\mathcal{Q}_{i+r}(t+dt) = ~~~~~~~~~~~~~~~~~~~~~~~~~~~~~~~~~~~~$$
\begin{align}\label{eq:mass_cur_update}   
   \begin{cases}
     \textbf{event}         & \textbf{prob.} \\
     m_i\mathcal{Q}_{i+r}+\lamt m_i(\xi_{i+r}m_{i+r}-\tilde{\xi}_{i+r+1}m_{i+r+1})\\+\lamt\left( \tilde{\xi}_{i+1}m_{i+1}-m_i+\xi_{i-1}m_{i-1}\right)\mathcal{Q}_{i+r} & dt \\
     m_i\mathcal{Q}_{i+r}-\lamt^2(\xi_im_i^2+\tilde{\xi}_{i+1}^2m_{i+1}^2) & \delta_{r,0}dt \\
     m_i\mathcal{Q}_{i+r} +\lamt^2(\tilde{\xi}_im_i^2+\xi_{i-1}^2m_{i-1}^2) & \delta_{r,-1}dt \\
     m_i\mathcal{Q}_{i+r} +\lamt^2\tilde{\xi}_{i+1}\xi_{i+1}m_{i+1}^2 & \delta_{r,1}dt \\
     m_i\mathcal{Q}_{i+r} -\lamt^2\xi_{i-1}\tilde{\xi}_{i-1}m_{i-1}^2 & \delta_{r,-2}dt \\
     m_i\mathcal{Q}_{i+r} & 1-\sum dt,
   \end{cases}
\end{align}
where $ \sum=1+\delta_{r,0}+\delta_{r,1}+\delta_{r,-1}+\delta_{r,-2}$ is the total exit rate. Using the above dynamical update rules, we obtain the following equation:
\begin{align}\label{eq:mq_tt}
    \frac{d}{dt}C_r^{m\cq}(t,t) = D\left(C_{r-1}^{m\cq}-2C_r^{m\cq}+C_{r+1}^{m\cq} \right)+A_r,
\end{align}
where $A_r$ is given by
\begin{align}
    \label{eq:Ar_1}\begin{aligned}
    A_r &= \frac{\lamt}{2}\langle m_im_{i+r}-m_im_{i+r+1}\rangle_c \\
    &-\frac{5\lamt^2}{6}\langle m_i^2\rangle\delta_{r,0}+\frac{\lamt^2}{6}\langle m_i^2\rangle\delta_{r,1} -\frac{\lamt^2}{6}\langle m_i^2\rangle\delta_{r,-2}.
    \end{aligned}
\end{align}
From the above equation, it is evident that $A_r$ can be represented in terms of the equal-time spatial correlation of masses, which can be calculated by using the following infinitesimal-time update rules,
$m_i(t+dt)m_{i+r}(t+dt) = $
\begin{align}\label{eq:mass_mass_update}   
   \begin{cases}
     \textbf{event}         & \textbf{prob.} \\
     \lamt\left( \tilde{\xi}_{i+1}m_{i+1}+\xi_{i-1}m_{i-1}\right)m_r\\
     +\lamt m_i(\tilde{\xi}_{i+r+1}m_{i+r+1}+\xi_{i+r-1}m_{i+r-1})\\
     (1-2\lamt) m_im_{i+r} & dt \\
     m_im_{i+r}+\lamt^2(m_i^2+\tilde{\xi}_{i+1}^2m_{i+1}^2+m_{i-1}^2\xi_{i-1}^2) & \delta_{r,0}dt \\
     m_im_{i+r} -\lamt^2(\tilde{\xi}_{i+1}m_{i+1}^2+\xi_{i}m_{i}^2) & \delta_{r,1}dt \\
     m_im_{i+r} +\lamt^2(\tilde{\xi}_{i}m_i^2+\xi_{i-1}m_{i-1}^2 & \delta_{r,-1}dt \\
     m_im_{i+r} -\lamt^2\xi_{i+1}\tilde{\xi}_{i+1}m_{i+1}^2 & \delta_{r,2}dt \\
     m_im_{i+r} -\lamt^2\xi_{i-1}\tilde{\xi}_{i-1}m_{i-1}^2 & \delta_{r,-2}dt \\
     m_im_{i+r} & 1-\sum dt,
   \end{cases}
\end{align}
where $ \sum=1+\delta_{r,0}+\delta_{r,1}+\delta_{r,-1}+\delta_{r,-2}+\delta_{r,2}$ is the total exit rate. Using the above dynamical update rules we obtain the following equation:
\begin{align}\label{eq:dmm_rtt_1}
    \frac{d}{dt}C_r^{mm}(t,t) = 2D\sum\limits_{k}\abr{m_0\Delta_{r,k}m_k}_c +B_r,
\end{align}
where the source term $B_r$ can be expressed in terms of the second moment of a single site mass as, 
\begin{align}
\begin{aligned}
    B_r = \frac{\lamt^2\abr{m_i^2}}{6}[10\delta_{r,0}-6(\delta_{r,1}+\delta_{r,-1})
    +(\delta_{r,2}+\delta_{r,-2})].
\end{aligned}
\end{align}
Then using the steady-state condition $d C_r^{mm}(t,t)/dt=0$, we obtain:
\begin{align}\label{eq:dmm_rtt}
    2D\left[C_{r+1}^{mm}-2C_r^{mm}+C_{r-1}^{mm}\right] +B_r=0.
\end{align}
Now, Eq.(\ref{eq:dmm_rtt}) can be exactly solved by employing a generating function method (which is slightly different than that used in \cite{das_spatial_2016}),
\begin{align}
    G(z) = \sum\limits_{r=0}^{\infty}C_r^{mm}z^{r};
\end{align}
we present it here for completeness of the various calculations and derivations of a set of fluctuation relations, which will follow in the subsequent sections.
By multiplying both sides of Eq.(\ref{eq:dmm_rtt}) by $z^r$ and then summing over $r$, we can express the equation in terms of the generating function,
\begin{widetext}
\begin{equation*}
    G(z) = \frac{C_0^{mm}[6-(10-6z+z^2)\lamt]-6zC_1^{mm}-(10-6z+z^2)\rho^2}{6(1-z)^2},
\end{equation*}
\end{widetext}
where we use identities, $\sum_{r=0}^{\infty}C_{r+1}^{mm}z^r=[G(z)-C_0^{mm}]/z$, $\sum_{r=0}^{\infty}C_{r-1}^{mm}z^r=C_1^{mm}+zG(z)$, and $\abr{m_i^2}=C_0^{mm}+\rho^2$. From the equation above, it is evident that $G(z)$ has a second-order pole singularity at $z=1$. However, considering that $G(z=1)$ represents the sum of density correlations and thus should be finite, implying that $\lim_{z\to 1}G(z)$ must also be finite. Therefore the numerator, and its derivative, must also vanish as $z\to 1$. From this root-cancellation conditions, we get the following two equations,
\begin{align}
    C_0^{mm}(6-5\lamt)-6C_1^{mm}-5\rho^2 =0,
\end{align}
and 
\begin{align}
    2\lamt C_0^{mm}+3C_1^{mm}-2\rho^2 =0.
\end{align}
By solving the above equations, we finally obtain the desired solution for the generating function,
\begin{align}
    G(z) = \frac{2\lamt\rho^2}{3-2\lamt} -\frac{\lamt\rho^2}{2(3-2\lamt)}z ,
\end{align}
which immediately leads to the explicit analytical expression for the steady-state spatial correlation function $C_r^{mm}$ for mass,
\begin{align}\label{eq:mass-mass}
    C_r^{mm}=\langle m_im_{i+r}\rangle-\rho^2=
    \begin{cases}
      \frac{2\lamt\rho^2}{3-2\lamt} & \text{for}~ r=0
      \\
      -\frac{\lamt\rho^2}{2(3-2\lamt)} & \text{for}~ |r|=1
      \\
    0 & \text{otherwise}.
    \end{cases}
\end{align} 
Note that the above correlation function was previously calculated through a different method in Ref.  \cite{das_spatial_2016}.
Now the steady-state spatial correlation function for mass can be readily expressed in terms of the particle mobility $\chi$,
\begin{align}\label{eq:mass-mass_f}
    C_r^{mm} = \frac{\chi}{\lamt}(4\delta_{r,0}-\delta_{r,1}-\delta_{r,-1}).
\end{align}
Now, by summing both sides of Eq.(\ref{eq:mass-mass_f}) over $r$, we obtain a fluctuation relation between mass fluctuation and the mobility (equivalently, the current fluctuation),
\begin{align}
    \sum\limits_{r=-\infty}^{\infty} C_r^{mm} = \frac{2\chi}{\lamt}.
\end{align}
Now, in the steady state, we write $A_r$ by simply using Eq.(\ref{eq:mass-mass_f}) in Eq.(\ref{eq:Ar_1}),
\begin{align}\label{eq:A_r}
    A_r = -\frac{5}{2}\chi
        (\delta_{r,0}-\delta_{r,-1})+\frac{1}{2}\chi
        (\delta_{r,1}-\delta_{r,-2}).
\end{align}
We now express Eq. (\ref{eq:mq_tt}) in the Fourier space as 
\begin{equation}\label{eq:dmq_tt}
\frac{d}{dt}\tilde{C}_q^{m\cq}(t,t)=-D\omega_q\tilde{C}_q^{m\cq}(t,t)+\tilde{f}_q(t),
\end{equation}
where the Fourier transform of the source term $A_r$ is expressed as $\tilde{f}_q(t^\prime)$ in the following equation:
\begin{equation}\label{eq:f_q}
\tilde{f}_q = -\chi(1-e^{-iq})\left(1+\frac{1}{2}\omega_q\right).
\end{equation}
Equation (\ref{eq:dmq_tt}) can now be integrated directly to obtain the following equation,
\begin{equation}\label{eq:mq_tt_f}
\tilde{C}_q^{m\cq}(t^\prime,t^\prime) = \int\limits_0^{t^\prime} dt^{\prime\prime} e^{-D\omega_q(t^\prime-t^{\prime\prime})}\tilde{f}_q(t^{\prime\prime}).
\end{equation}
The above equation describes the equal-time dynamic correlation of mass-current, which appears in Eq.(\ref{eq:crQQtt}) and Eq.(\ref{eq:mq_qt}), making it necessary for the calculation of the equal-time current-current correlation and unequal-time mass-current, respectively. 
By substituting Eq.(\ref{eq:mq_tt_f}) into Eq.(\ref{eq:mq_qt}), leading to the following expression,
\begin{equation}\label{eq:mq_ttp_f}
\tilde{C}_q^{m\cq}(t,t^\prime) = \int\limits_0^{t^\prime} dt^{\prime\prime} e^{-D\omega_q(t-t^{\prime\prime})}\tilde{f}_q(t^{\prime\prime}).
\end{equation}

\subsection{Time-integrated current fluctuation}\label{subsec:QQT}

In this section, we apply the theoretical framework established in the previous section to finally compute the time-integrated bond-current fluctuation for MCM I. To achieve this, we insert Eq. (\ref{eq:mq_tt_f}) into Eq. (\ref{eq:crQQtt}), yielding an explicit expression for the time-integrated bond-current fluctuation at equal times, as follows:
\begin{equation}\label{eq:crQQtt_f}
\begin{aligned}
&C_r^{\cq\cq}(t,t) = \int\limits_{0}^{t}dt^\prime \Gamma_r(t^\prime) +\\ &\frac{D}{L}\sum\limits_{q}\int\limits_{0}^{t}dt^\prime \int\limits_0^{t^\prime} dt^{\prime\prime} e^{-D\omega_q(t^\prime-t^{\prime\prime})}\tilde{f}_q(t^{\prime\prime})(1-e^{iq})[2-\omega_{qr}].
\end{aligned}
\end{equation}
Furthermore, by plugging in the equal-time current-current dynamic correlation $C_r^{\cq\cq}(t,t)$ from Eq.(\ref{eq:crQQtt_f}) and the unequal-time mass-current $\tilde{C}_q^{m\cq}(t,t^\prime)$ from Eq.(\ref{eq:mq_ttp_f}) into Eq.(\ref{eq:qq_qt}), we can obtain the final expression for the current-current dynamic correlation at unequal time:
\begin{widetext}
  \begin{equation}
    \label{eq:QQ_ttp_r}
  \begin{aligned}
   C^{\cq\cq}_r(t,t^\prime) =
   & t^\prime \Gamma_r - 
    \frac{\chi D}{L}  \sum\limits_{q}
     \int\limits_{0}^{t^\prime} \dd{t^{\prime \prime}}
     \int\limits_{0}^{t^{\prime \prime}} 
     \dd{t^{\prime \prime \prime}} e^{-D\omega_q(t^{\prime \prime} - t^{\prime
     \prime \prime})} 
    \omega_q (1+\frac{1}{2}\omega_q) (2-\omega_{qr})
    \\
   &- \frac{\chi D}{L}  \sum\limits_{q}
     \int\limits_{t^\prime}^{t} \dd{t^{\prime \prime}}
     \int\limits_{0}^{t^{\prime}} 
     \dd{t^{\prime \prime \prime}} e^{-D\omega_q(t^{\prime \prime} - t^{\prime
     \prime \prime})} 
    \omega_q (1+\frac{1}{2}\omega_q) e^{-iqr}.
\end{aligned}
\end{equation}
\end{widetext}

We obtain the time-integrated bond-current fluctuation $\langle \mathcal{Q}_i^2(T) \rangle = C_0^{\mathcal{Q} \mathcal{Q}}(T,T)$ from Eq.(\ref{eq:QQ_ttp_r}), by putting $t' =t=T$ and $r=0$,
\begin{equation}\label{eq:QQ_t_0}
    \langle \mathcal{Q}_i^2(T)\rangle = \frac{2\chi T}{L} +\frac{2\chi}{L}\sum\limits_{n=1}^{L-1}\left (1 + \frac{\omega_n}{2}\right ) \frac{(1-e^{-D\omega_nT})}{D\omega_n}, 
\end{equation}
where  $\omega_n=2(1-\cos(2\pi n/L))$, with $n=0,1,\cdots, L-1$. 
If we take the $T \to \infty$ limit first (i.e., $T \gg L^2$), we immediately obtain
\begin{align}\label{eq:asyyqq}
  \langle \mathcal{Q}_i^2(T)\rangle \simeq \frac{2\chi T}{L}+ \frac{\chi}{D}\left(\frac{L}{6}-\frac{1}{L} \right)
   = \frac{2\chi T}{L} \left[ 1 + {\cal O}\left( \frac{L^2}{DT} \right) \right].
\end{align}
In Eq. (\ref{eq:QQ_t_0}), we have identified two distinct time regimes that correspond to two of the following cases.
\\
\begin{center}{Case 1: $DT \ll 1$}\end{center}
In the limit $DT \ll 1$, the system does not have sufficient time for building up spatial correlations at neighboring sites, resulting in the bond-current fluctuation having no information of the spatial structure.
In equation (\ref{eq:QQ_t_0}), we expand the exponential up to linear order for $DT\ll1$ and obtain
\begin{align}
    \langle \mathcal{Q}_i^2(T)\rangle = \frac{2\chi T}{L} +\frac{2\chi}{L}\sum\limits_{n=1}^{L-1}\left (1 + \frac{\omega_n}{2}\right )T.
\end{align}
To further simplify the above equation, we utilize the identity $\sum_{n=1}^{L-1}\omega_n=2L$, resulting in 
\begin{equation}\label{eq:gamma_0t}
     \langle \mathcal{Q}_i^2(T)\rangle=\Gamma_0 T = 4\chi T,
\end{equation}
where $\Gamma_0$ is the strength of fluctuating current as mentioned in Eq.(\ref{eq:gamma_r_f}) and later calculated exactly.

\begin{figure}[H]
    \centering
    \includegraphics[scale=0.5]{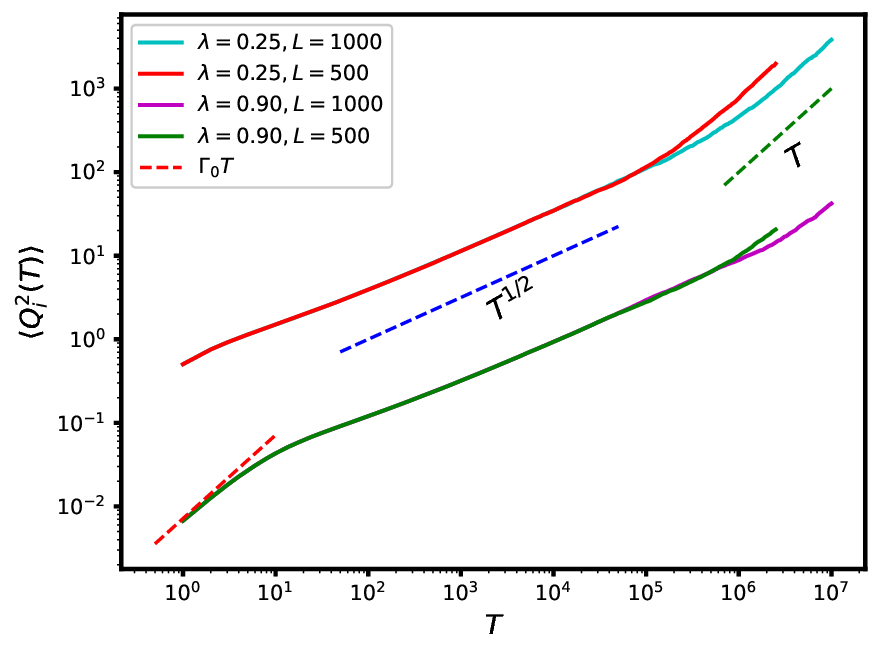}
    \caption{The time-integrated bond-current fluctuation, $\langle \mathcal{Q}_i^2(T) \rangle$, plotted as a function of time $T$ for various chipping parameter and system sizes. The cyan ($\lambda=0.25$, $L=1000$), magenta ($\lambda=0.90$, $L=1000$), red ($\lambda=0.25$, $L=500$), and green ($\lambda=0.90$, $L=500$) lines lines represent simulation data for global density $\rho=1$. The red dashed line represents the behavior $\Gamma_0T$ for $\lambda = 0.90$ as mentioned in Eq.(\ref{eq:gamma_0t}), while the blue and green dashed lines represent sub-diffusive $\sim T^{1/2}$ and diffusive $\sim T$ growth, respectively, as mentioned in Eq.(\ref{eq:qi2t_tt_scalling_full}).}
    \label{fig:Qi2t_Mcm1_unscale}
\end{figure}

In Figure \ref{fig:Qi2t_Mcm1_unscale}, we present simulation data for time-integrated bond-current fluctuation $\langle \mathcal{Q}_i^2(T) \rangle$ as a function of time $T$. The plot reveals three distinct behaviors of temporal growth of the current fluctuations: Short-time $T$, intermediate-time $T^{1/2}$, and long-time $T$ growths. We have examined the effects of chipping by varying the values of $\lambda =0.25$ and $0.90$ for two different system sizes $L=500$ and $L=1000$. Also, we provide two plots concerning current fluctuations that overlap in the region $DT \ll 1$ for two system sizes with $\lambda$ kept fixed.

\begin{figure}[H]
    \centering
    \includegraphics[scale=0.5]{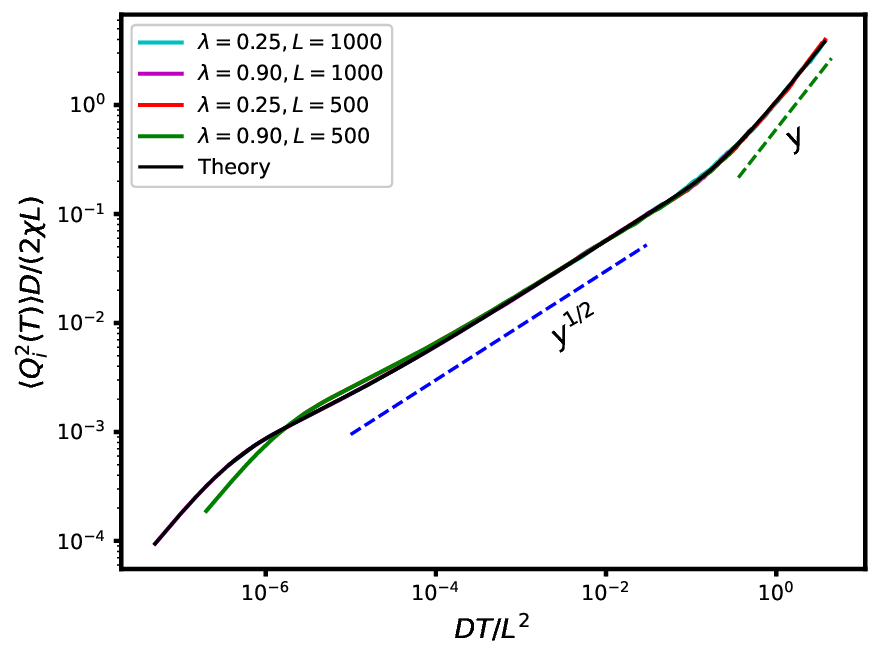}
    \caption{Scaled time-integrated bond-current fluctuation, $\langle \mathcal{Q}_i^2(T) \rangle D/(2\chi L)$, plotted against scaled time $DT/L^2$ for different chipping parameters and system sizes at a global density of $\rho=1$. The cyan ($\lambda=0.25$, $L=1000$), magenta ($\lambda=0.90$, $L=1000$), red ($\lambda=0.25$, $L=500$), and green ($\lambda=0.90$, $L=500$) lines lines represent simulation data. The two guiding dashed lines depict sub-diffusive behavior as $\sim y^{1/2}$ (blue) at early times, followed by a diffusive growth as $\sim y$ (green) at longer times. These trends are based on the scaling function $\mathcal{W}(y)$ [as in Eq.(\ref{eq:qi2t_tt_scalling})]. The black solid line corresponds to theoretical results [as in Eq.(\ref{Wy})] and demonstrates excellent agreement with the simulations.}
    \label{fig:qi2t_diff_lam}
\end{figure}

\begin{center}{Case 2: $DT \gg 1$}\end{center}
In the limit $DT \gg 1$, spatial correlations build up in the system. Interestingly, Eq.(\ref{eq:asyyqq}) suggests that, in the large-time limit, $\langle Q_i^2(T) \rangle$ asymptotically approaches $2\chi T/L$, indicating that, in the long-time regime, there are only three macroscopic parameters, which are relevant and thus can be used to characterize the diffusive growth: the bulk-diffusion coefficient $D$, the mobility $\chi$, and system size $L$. Indeed, one would expect that $\langle \mathcal{Q}_i^2\rangle$ and $T$ should be related through a precise scaling combination involving the above mentioned macroscopic quantities as given in Eq.(\ref{eq:scalling_cur_1}). To further analyze the behavior, we introduce a scaled time-integrated bond-current fluctuation as follows:
\begin{equation}\label{eq:scalling_cur_1}
\begin{aligned}
&\frac{D\langle \mathcal{Q}_i^2(T)\rangle}{2\chi L}  \equiv  \mathcal{W} \left(\frac{DT}{L^2}\right) 
  \\
    &=\lim\limits_{L\to \infty} \frac{DT}{L^2} + \lim\limits_{L\to \infty} \frac{1}{L^2}\sum\limits_{q}\left(1+\frac{\omega_q}{2}\right)\frac{1-e^{-\omega_qDT}}{\omega_q}.
    \end{aligned}
\end{equation}
Or, equivalently, we can exactly write the scaling function as
\begin{eqnarray}
  \mathcal{W}(y) = \lim\limits_{L\to \infty} \left[  y + \frac{1}{L^2}\sum\limits_{q}\left(1+\frac{\omega_q}{2}\right)\frac{1-e^{-\omega_q L^2 y}}{\omega_q} \right],~~
  \label{Wy0}
\\
= \lim \limits_{L\to \infty} \left[  y + \frac{1}{L^2}\sum\limits_{q}  \frac{1-e^{-\omega_q L^2 y}}{\omega_q} \right] + o\left( \frac{1}{L} \right),~~
\label{Wy}
\end{eqnarray}
where $\omega_q=2(1-\cos q)$, with $q=2\pi n/L$ and $n=1,2,\cdots, L-1$. In the expression as given in the right hand side of  Eq. \ref{Wy0}, the scaling function $\mathcal{W}(y)$ can be approximately written in terms of the following integral representation,
\begin{align}
\begin{aligned}\label{eq:w_y1}
    \mathcal{W}(y)\simeq y+&\lim\limits_{L\to \infty}\frac{1}{\pi L}\int\limits_{\frac{2\pi}{L}}^{\pi} dq \frac{\left ( 1- e^{-L^2 y \omega(q)}\right)}{\omega(q)}\\&+\lim\limits_{L\to \infty}\frac{1}{2\pi L}\int\limits_{\frac{2\pi}{L}}^{\pi} dq \left ( 1- e^{-L^2 y \omega(q)}\right).\\
    \end{aligned}
\end{align}
By using variable transformation $z = \omega(q)L^2$  in  Eq.(\ref{eq:w_y1}) and taking the infinite system-size limit $L\to\infty$, we obtain the following expression,
\begin{align}
    \mathcal{W}(y) \simeq y+\frac{1}{2\pi}\int\limits_{4\pi^2}^{\infty}dz  \frac{\left(1-e^{-zy}\right)}{z^{\frac{3}{2}}},
\end{align}
where we have used that the third term in the rhs of eq. (\ref{eq:w_y1}) gives a subleading contribution, which vanishes in the scaling limit. 
After performing the integration, we finally write
\begin{equation}\label{eq:Scalling_cur}
\begin{aligned}
    \mathcal{W}(y) \simeq y+\sqrt{\frac{y}{\pi}}\text{erfc}(2\pi\sqrt{y})+\frac{1-e^{-4\pi^2y}}{4\pi^2},
    \end{aligned}
\end{equation}
where $\text{erfc}(z)=1-\text{erf}(z)$, with error function,
\begin{align}
    \text{erf}(z) = \frac{2}{\sqrt{\pi}}\int\limits_{0}^{z}e^{-t^2}dt.
\end{align}
From Eq.(\ref{eq:Scalling_cur}), we can now determine the asymptotic form of  $\mathcal{W}(y)$ in the following two regimes of $y$,
\begin{equation}\label{eq:qi2t_tt_scalling}
\mathcal{W}(y) \simeq
\begin{cases}
  \sqrt{\frac{y}{\pi}} & \text{for}~  y \ll 1
  \\
   y & \text{for} ~y\gg 1.
\end{cases}
\end{equation}
In Fig.\ref{fig:qi2t_diff_lam}, we plot the scaled time-integrated bond-current fluctuation $\langle \mathcal{Q}_i^2(T) \rangle D/(2\chi L)$ as a function of scaled time $DT/L^2$ for different chipping parameters and system sizes for a global density $\rho=1$. The colored lines are obtained from simulations, and the black solid line corresponds to theory as in Eq.(\ref{Wy}). Two guiding (dashed) lines, which represent sub-diffusive  $\sim y^{1/2}$ behavior (blue) at early (but still large) times, followed by diffusive growth as $\sim y$ (green) at longer times, as given in Eq.(\ref{eq:qi2t_tt_scalling}).

Now combining all three above mentioned temporal regimes, we summarize the asymptotic behaviors of the time-integrated bond-current $\langle \mathcal{Q}_i^2(T) \rangle$ as following:
\begin{equation}\label{eq:qi2t_tt_scalling_full}
\langle\mathcal{Q}_i^2(T)\rangle=
\begin{cases}
   4\chi T & \text{for}~ DT\ll  1\\
   \frac{2\chi}{\sqrt{D\pi}}T^{\frac{1}{2}} & \text{for}~ 1 \ll DT \ll L^2 \\
   \frac{2\chi T}{L} & \text{for} ~DT\gg L^2,
\end{cases}
\end{equation}
where the second and third regimes, as shown above, are related through a single scaling function ${\cal W}(y)$. 

\subsubsection{Space-time integrated current fluctuations}

In this section, we focus on the variance of steady-state $\langle\mathcal{Q}_{sub}^2(l,T)\rangle-\langle\mathcal{Q}_{sub}(l,T)\rangle^2=\langle \mathcal{Q}_{sub}^2(l,T) \rangle$ of the cumulative (space-time integrated) particle current $\cq_{sub}(l,T)=\sum_{i=0}^{l-1}\cq_i(T)$ across a subsystem of size $l$ and up to time $T$. The subsystem current fluctuation $\langle\mathcal{Q}_{sub}^2(l,T)\rangle$ can be written as 
\begin{equation}\label{eq:QQ_vt}
\begin{aligned}
 \langle \mathcal{Q}_{sub}^2(l,T) \rangle
     = \abr{\sum\limits_{i=1}^{l-1}\cq_i(T)\sum\limits_{j=1}^{l-1}\cq_j(T)}.
\end{aligned}
\end{equation}
Clearly, the sum on the right-hand side of the equation can be expressed in terms of the current-current dynamic correlation at equal times as
\begin{align}
    \langle \mathcal{Q}_{sub}^2(l,T) \rangle = lC_0^{\mathcal{Q}\mathcal{Q}}(T,T)+\sum\limits_{r=1}^{l-1}2(l-r)C_r^{\mathcal{Q}\mathcal{Q}}(T,T).
\end{align}
Now, by using Eq.(\ref{eq:crQQtt_f}) in the above equation and then employing the following identity,
\begin{align}
    \sum\limits_{r=1}^{l-1}2(l-r)(2-\omega_{rq})=2\left(\frac{\omega_{lq}-l\omega_{q}}{\omega_{q}}\right),
\end{align}
we obtain, after some algebraic manipulations, the following expression,
\begin{equation}\label{eq:QQ_vt_f}
  \begin{aligned}
   \langle \mathcal{Q}_{sub}^2(l,T) \rangle & =
    2\chi l T + 2\chi T (1-\delta_{l,L})  
    - \frac{2\chi D}{L} \\
    & \times \sum\limits_{q}\frac{(D\omega_qT-1+e^{-\omega_qDT})}{(\omega_q D)^2}\big( 1+\frac{1}{2}\omega_q\big)\omega_{ql},    
\end{aligned}
\end{equation}
where $\omega_q=2(1-\cos q)$ with $q=2\pi n/L$ and $n=1,2,\cdots, L-1$. The subsystem size $l$ comes into play through the Fourier mode $\omega_{ql}$ alone. We now derive the asymptotic dependence of Eq.(\ref{eq:QQ_vt_f}) on the subsystem size $l$ and time $T$, first by considering the limit $T\gg 1$ followed by $l\gg 1$, and then by reversing the order of the limits, i.e., $l\gg 1$ followed by $T\gg 1$.
\\\\
\begin{center}{Case 1: $T \gg 1$ and $l\gg 1$}\end{center}
\vspace{0.25cm}
We first consider the limit $T\gg 1$ followed by $l\gg 1$; in that case, the above Eq.(\ref{eq:QQ_vt_f}) simplifies to
\begin{equation}  
\begin{aligned}
    &\langle \mathcal{Q}_{sub}^2(l,T) \rangle\\ &\simeq \frac{2\chi l^2T}{L}+\frac{2\chi D}{L}
    \sum\limits_{q}\frac{(1-e^{-D\omega_qT})}{D^2\omega_q^2}\left(1+\frac{\omega_q}{2}\right)\omega_{ql}.
    \end{aligned}
\end{equation}
Now, in the limit of $L\to\infty$, the sum in the above equation can be approximated as an integral,
\begin{align}
    &\langle \mathcal{Q}_{sub}^2(l,T) \rangle\\ &\simeq \frac{2\chi D}{\pi }\int\limits_{0}^{\pi} dq \frac{(1-e^{-D\omega (q)T})}{D^2\omega (q)^2}\left(1+\frac{\omega(q)}{2}\right)\omega (ql).
\end{align}
By using the approximation $\omega(lq)\simeq l^2q^2$ for a finite subsystem size $l$, and then using a variable transformation $z=DTq^2$, we obtain the following form,
\begin{align}
    \langle \mathcal{Q}_{sub}^2(l,T) \rangle \simeq \frac{\chi l^2\sqrt{T}}{\pi\sqrt{D}}\int\limits_{0}^{\infty} dz (1-e^{-z})z^{-3/2},
\end{align}
which, by using the identity $\int_{0}^{\infty}dz (1-e^{-z})z^{-3/2}=2\sqrt{\pi}$, boils down to the asymptotic form,
\begin{align}
    \frac{\langle \mathcal{Q}_{sub}^2(l,T) \rangle}{lT}\simeq \frac{2\chi}{\sqrt{\pi D}}\frac{l}{\sqrt{T}}.
\end{align}
\\\\
\begin{center}{Case 2: $l \gg 1$ and $T\gg 1$}\end{center}
\vspace{0.25cm}
In this specific order of limits (i.e., $l \gg 1$ limit first and then $T \gg 1$), the equation (\ref{eq:QQ_vt_f}) can be expressed in integral form,
\begin{align}
\begin{aligned}
    &\frac{\langle \mathcal{Q}_{sub}^2(l,T) \rangle}{lT}\\
    &\simeq 2\chi +\frac{2\chi}{l}
    +\frac{4\chi D}{lT\pi}\int\limits_{0}^{\pi} dq \frac{[D\omega(q)T-1+e^{-\omega(q)DT}]}{D^2\omega(q)^2},
\end{aligned}   
\end{align}
 where we use an approximation $\omega(ql) \simeq 2$.
Again, by using the variable transformation $z=DTq^2$, we obtain
\begin{align}
\begin{aligned}
    &\frac{\langle \mathcal{Q}_{sub}^2(l,T) \rangle}{lT}\\
    &\simeq 2\chi +\frac{2\chi}{l}-\frac{2\chi\sqrt{DT}}{\pi l}\int\limits_{0}^{\infty} dz (z-1+e^{-z})z^{-\frac{5}{2}}.
    \end{aligned}
\end{align}
By using the identity $\int_{0}^{\infty} dz (z-1+e^{-z})z^{-5/2}=4\sqrt{\pi}/3$ in the above equation, we find the following asymptotic behavior, 
\begin{align}
    \frac{\langle \mathcal{Q}_{sub}^2(l,T) \rangle}{lT} \simeq 2\chi-\frac{8\chi}{3}\frac{\sqrt{D}}{\pi}\frac{\sqrt{T}}{l}
\end{align}
Hence, the asymptotic expression for the variance of the cumulative subsystem current, as given in eq. (\ref{eq:QQ_vt_f}), in fact depends on the order of limits for $T\gg 1$ and $l\gg1$, i.e.,
\begin{equation}\label{eq:asy_Q_VT}
\frac{\langle \mathcal{Q}_{sub}^2(l,T) \rangle}{lT} \simeq
\begin{cases}
\frac{2\chi}{\sqrt{\pi D}}\frac{l}{\sqrt{T}}~ & \text{for}~T \gg 1, l \gg 1 , \\
2\chi-\frac{8\chi}{3}\frac{\sqrt{D}}{\pi}\frac{\sqrt{T}}{l}~ & \text{for}~ l \gg 1, T \gg 1.
\end{cases}
\end{equation}
The first expression in the equation above results from taking the limits in the following sequence: first $T\gg 1$ and then $l\gg 1$. In this specific order of limits, the scaled function $\langle \mathcal{Q}_{sub}^2(l,T) \rangle/lT$ decreases as $1/\sqrt{T}$ and eventually diminishes as $T$ approaches infinity. On the other hand, if we reverse the order of limits, i.e., first $l\gg 1$ limit and $T\gg 1$ limit, we derive the second asymptotic expression as in Eq.(\ref{eq:asy_Q_VT}). Notably, when infinite subsystem-size limit $l \to \infty$ is taken first, the scaled fluctuation of the subsystem current $\langle \mathcal{Q}_{sub}^2(l,T) \rangle/lT$ converges to a finite number - twice the mobility $2\chi$ - as $T$ increases,
\begin{equation}\label{eq:sig_Q}
    \sigma_{\mathcal{Q}}^2 \equiv \lim\limits_{T\to\infty} \left[ \lim \limits_{l\to\infty} \frac{\langle \mathcal{Q}_{sub}^2(l,T) \rangle}{lT} \right] = \sum\limits_{r} \Gamma_r.
\end{equation}
Indeed the fluctuation relation as in Eq. (\ref{eq:sig_Q}) could be viewed as a nonequilibrium version of the Green-Kubo relation, which is well known for equilibrium systems.
If we consider $l=L\gg 1$ (corresponding to the case when the bond current is summed over the entire system), we can recast the above fluctuation relation as
\begin{equation}
\label{eq:large_L_Q}
    \lim_{L\to\infty}\frac{\langle \mathcal{Q}_{sub}^2(L,T) \rangle}{LT} = 2\chi = \sum_{r}\Gamma_r.
\end{equation}
Interestingly, the above expression is valid for any finite time $T$. This is due to the fact that, by definition, the diffusive part of the total current vanishes over the full system size, i.e., $\sum_{i=1}^{L}\mathcal{J}^{(d)}_i=0$. As a result, we are left with the space-time current having only the contribution from the fluctuating part, thus leading to the sum rule $\sum_r\Gamma_r = 2 \chi$ as given in eq. \eqref{eq:large_L_Q}.

\begin{figure}[H]
    \centering
    \includegraphics[scale=0.5]{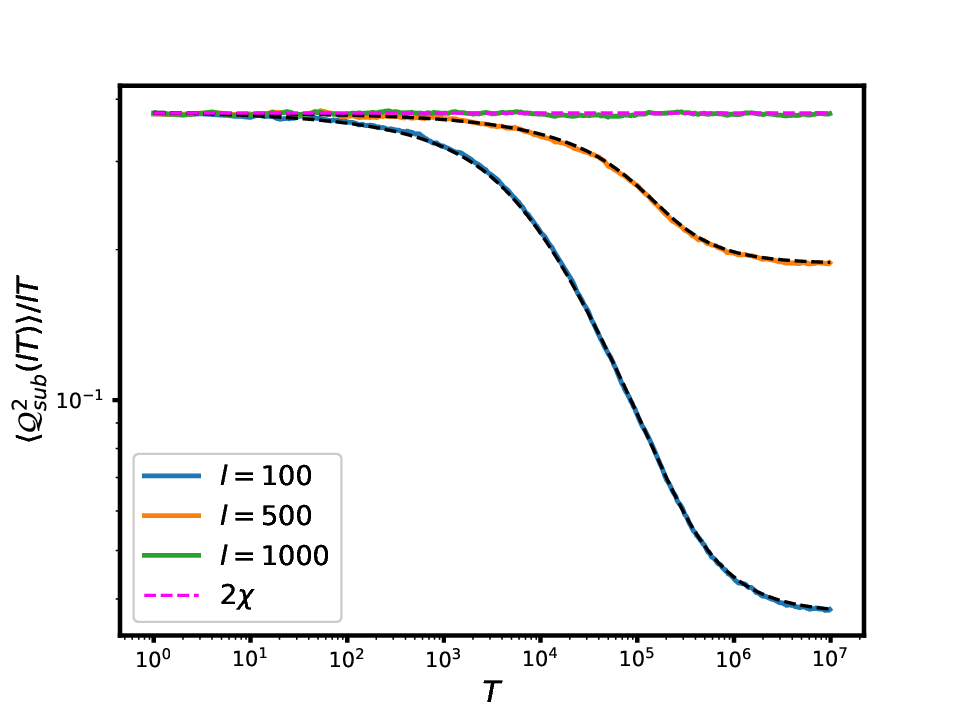}
    \caption{The scaled space-time-integrated bond-current fluctuation, $\langle \mathcal{Q}_{sub}^2(l,T) \rangle$, is displayed as a function of time $T$ for various sub-system sizes: $l=100$ (lower blue solid line), $500$ (middle orange solid line), and $1000$ (top solid green line). The chosen chipping parameter for the model is $\lambda=0.25$ (fixed), with a system size of $L=1000$ and a global density of $\rho=1$. The black dashed line in the plot corresponds to the theoretical prediction derived from Eq.(\ref{eq:QQ_vt_f}), and it precisely aligns with the respectable simulation data. Moreover, when the subsystem size equals the full system size, i.e., $l=L$, $\langle\mathcal{Q}_{sub}^2(L,T)\rangle/LT$ follows the behavior of $2\chi$ (magenta dashed lines), as indicated by Eq.(\ref{eq:large_L_Q}).}
    \label{fig:Qvt_L1000}
\end{figure}
In Figure \ref{fig:Qvt_L1000}, we plot the scaled subsystem current fluctuation $\langle\mathcal{Q}_{sub}^2(l,T)\rangle/lT$ against time $T$ for various subsystem sizes: $l=100$ (blue line), $l=500$ (orange line), and $l=L=1000$ (green line). The black dashed lines represent theoretical predictions that closely match the simulation data. A magenta dotted line at $2\chi$ overlays the data when $l=L$, indicating a limit where $l\to\infty$ is taken first. Notably, for smaller subsystem sizes, i.e., when  $T\to\infty$ limit is taken first, $\langle\mathcal{Q}_{sub}^2(l,T)\rangle/lT$ tends to zero.

\subsection{Instantaneous bond-current fluctuations}

\label{Sec:Pospeck_intro}

In this section, we calculate the spatio-temporal correlation function $C^{\mathcal{J} \mathcal{J}}_r(t) = \langle \mathcal{J}_i(t) \mathcal{J}_{i+r}(0) \rangle_c$ of the instantaneous bond current $\mathcal{J}_i(t)$ from the already calculated correlation function involving time-integrated bond current and show that the instantaneous bond current is negatively correlated in time.
 This is accomplished by taking a double derivative of the time-integrated bond current correlation as following, 
\begin{align}
    C_r^{\cj\cj}(t,t^\prime) = \left[\frac{d}{dt}\frac{d}{dt^\prime}C_r^{\cq\cq}(t,t^\prime) \right].
\end{align}
Here, with the understanding that $t \geq t^\prime$, we proceed to differentiate Eq.(\ref{eq:QQ_ttp_r}) twice with respect to time, resulting in the expression
\begin{align}\label{eq:cJJ_rttp}
    \begin{aligned}
        C_r^{\cj\cj}(t,t^\prime) &= \Gamma_r\delta(t-t^\prime)\\
        &- \frac{\chi D}{L}\sum\limits_{q}
    e^{-D\omega_q(t - t^\prime)}\omega_q \left(1+\frac{\omega_q}{2}\right) e^{-iqr}.
    \end{aligned}
\end{align}
To investigate the temporal behavior of instantaneous bond current, we set $r=0$ and $t>t^\prime=0$ in Eq.(\ref{eq:cJJ_rttp}) and simplify the expression, to obtain the integral form of the bond current correlation as
\begin{align}
    C_0^{\cj\cj}(t,0)\simeq \frac{\chi D}{\pi}\int\limits_{0}^{\pi} dq e^{-D\omega(q)t}\omega(q)\left[1+\frac{\omega(q)}{2}\right],
\end{align}
where we take the thermodynamic limit $L\to\infty$ and $\omega(q)=2(1-\cos q)$. Now, by approximating $\omega(q)\simeq q^2$ and making a variable transformation $z=Dq^2t$, we  rewrite the above equation as 
\begin{align}\label{eq:c_cj_t_0}
    C_0^{\cj\cj}(t,0)\simeq  -\frac{\chi t^{-\frac{3}{2}}}{2\pi\sqrt{D}}\int\limits_{0}^{\infty}z^{\frac{1}{2}}e^{-z}dz,
\end{align}
where we have ignored the subleading term $O(t^{-5/2})$. We note that the sign of the dynamic correlation function for bond curent is negative as given in Eq.(\ref{eq:c_cj_t_0}). Finally, by using the integral $\int_{0}^{\infty}z^{1/2}e^{-z}dz=\sqrt{\pi}/2$, the asymptotic form of the dynamic current correlation function can be expressed in following form,
\begin{align}
  \label{power-law1}
    C_0^{\cj\cj}(t,0)\simeq \Gamma_0\delta(t) -\frac{\chi}{4\sqrt{\pi D}}t^{-\frac{3}{2}}.
\end{align} 
Notably, the negative (second) part in the right hand side of the above equation exhibits long-range power-law behavior, solely due to the contribution due to the dynamic correlation involving the diffusive current, i.e., $C_0^{\cj^{(d)}\cj^{(d)}}(t,0)\sim t^{-3/2}$. On the other hand, the fluctuating current is short-ranged $C_r^{\cj^{(fl)}\cj^{(fl)}}(t,0)=\delta(t)\Gamma_r$, which is delta-correlated in time and where $\Gamma_r$ represents the space-dependent strength of the fluctuating current. This particular behavior should be contrasted with that for the symmetric simple exclusion processes \cite{Sadhu_2016}, where, due to the lack of spatial correlations, the strength does not depend on space variable $r$ (the temporal part is delta-correlated as in the mass chipping models).

Now, to obtain the spatial correlation of instantaneous current, we calculated the two-point correlation function calculated at the same time, $t=t^\prime$, but at different space points separated by a distance $r$. In that case, Eq.(\ref{eq:cJJ_rttp}) can be expressed as:
\begin{align}\label{eq:cJJ_rtt}
    \begin{aligned}
        C_r^{\cj\cj}(t,t) = \Gamma_r\delta(0)- \frac{\chi D}{L}\sum\limits_{q}\omega_q \left(1+\frac{\omega_q}{2}\right) e^{-iqr}.
    \end{aligned}
\end{align}
After some algebraic manipulations, the above equation can be expressed in terms of the steady-state $(t\to\infty)$ density correlation $C_r^{mm}$,
\begin{align}\label{eq:cJJ_cr}
    \begin{aligned}
        C_r^{\cj\cj} &= \Gamma_r\delta(0)-D^2[3C_r^{mm}-(C_{r-1}^{mm}+C_{r+1}^{mm})].
    \end{aligned}
\end{align}
In the above equation, we can see that the spatial length scale associated with the instantaneous current is intimately connected to the space-dependent strength $\Gamma_r$ of the instantaneous current. There is an interesting observation: As $\Gamma_r$ directly influences the density correlation $C_r^{mm}$, we find that, in the steady state, the spatial extent over which the instantaneous bond currents are correlated  is inherently short-ranged and, moreover, is characterized by the same length scale, which governs the spatial density correlations. In contrast to the dynamic correlation of the fluctuating current, which constitutes two distinct components - (i) the fluctuating current, which is short-ranged both in terms of spatial and temporal dependencies, and (ii) the diffusive current, which is characterized by its short-ranged spatial behavior, but the long-ranged (power law) temporal behavior ($\sim t^{-3/2}$).

\begin{figure}[H]
    \centering
    \includegraphics[scale=0.5]{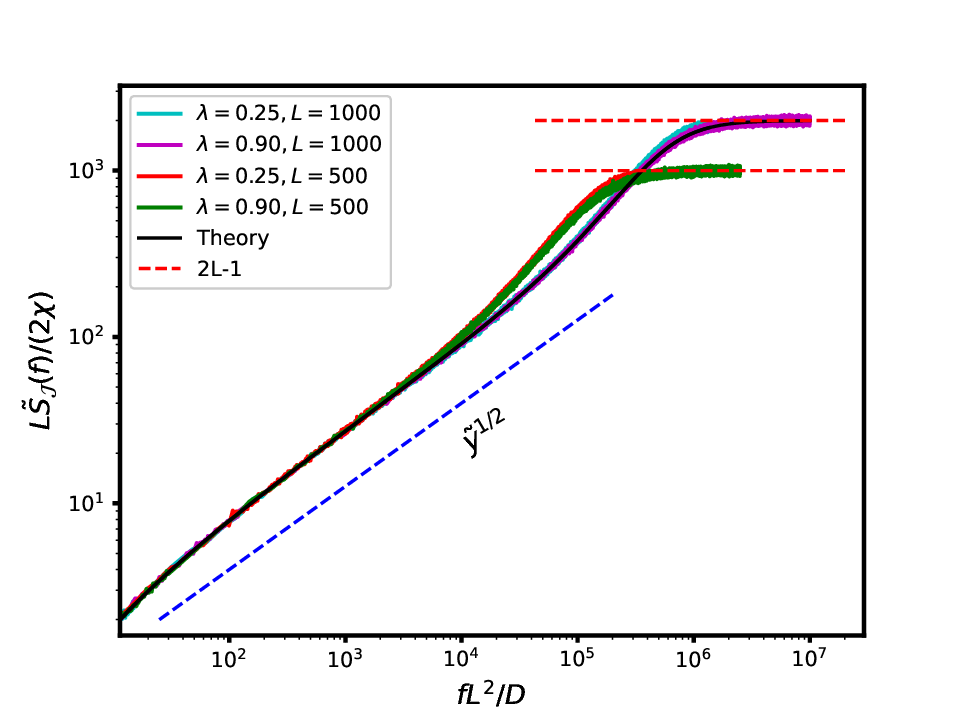}
    \caption{The scaled power spectrum of instantaneous
    currents, $L\tilde{S}_{\cj}(f)/(2\chi)$, is plotted as a function
    of scaled frequency $L^2f/D$ for various chipping parameters and system sizes.
    The cyan solid line corresponds to $\lambda=0.25$ and $L=1000$, the magenta solid line corresponds to $\lambda=0.90$ and $L=1000$, the red solid line corresponds to $\lambda=0.25$ and $L=500$, and the green solid line corresponds to $\lambda=0.90$ and $L=500$, all at a global density of $\rho=1$.
    The blue dashed line shows $\tilde{y}^{1/2}$ scaling behavior in the low-frequency regime as in Eq.(\ref{eq:assy_sj}) and red dashed lines represent $L\tilde{S}_{\cj}(f)/(2\chi)$ diverges as system size $2L-1$ at the high-frequency limit. The solid color lines represent the simulation results, while the black solid line represents the theoretical predictions of Eq.(\ref{eq:current_power_speck_scale}), which perfectly matches the simulation data.}
    \label{fig:Power_speck_cur_MCM_I}
    \end{figure}

We also study the power spectrum for the instantaneous bond  current $\cj_i$, which, by using the \textit{ Wiener-Khinchin theorem} \cite{MacDonald}, can be expressed as 
\begin{equation}\label{eq:ft_def}
S_{\cj}(f) = \int\limits_{-\infty}^{\infty} dt  C_0^{\cj\cj}(t,0)e^{2\pi i f t}.
\end{equation}
By putting $r=0$ and $t^\prime=0$ in Eq.\eqref{eq:cJJ_rttp}, we integrate the right hand side of the above equation and obtain 
\begin{equation}\label{eq:power_speck_icur}
S_{\cj}(f) = \frac{2\chi(\rho)}{L}+\frac{2\chi(\rho)}{L}\sum\limits_{q}\big(1+\frac{1}{2}\omega_q\big)\frac{4\pi^2f^2}{D^2\omega_q^2+4\pi^2f^2}.
\end{equation}
Here $\omega_q=2(1-\cos q)$ with $q=2\pi n/L$ and $n=1,2,\cdots, L-1$.
By subtracting the zero-th mode ($f=0$), we rewrite the above expression through a suitably defined power spectrum  $\tilde{S}_\cj(f)=S_\cj(f)-S_\cj(0)$,
\begin{equation}\label{eq:current_power_speck}
\tilde{S}_\cj(f)=\frac{2\chi(\rho)}{L}\sum\limits_{q}\big(1+\frac{1}{2}\omega_q\big)\frac{4\pi^2f^2}{D^2\omega_q^2+4\pi^2f^2}.
\end{equation}
Now we rescale frequency as $\tilde{y}=fL^2/D$ and introduce a scaling function $\mathcal{H}$, that relates to $\tilde{S}_\cj(f)$ as
\begin{equation}\label{eq:current_power_speck_scale}
\begin{aligned}
\mathcal{H}\left(\frac{L^2f}{D}\right)&=\frac{L\tilde{S}_\cj(f)}{2\chi(\rho)}\\
&=\lim_{L\to\infty}\sum\limits_{q}\big(1+\frac{1}{2}\omega_q\big)\frac{4\pi^2\left(\frac{L^2f}{D}\right)^2}{L^4\omega_q^2+4\pi^2\left(\frac{L^2f}{D}\right)^2}.
\end{aligned}
\end{equation}
The above expression can be represented in an integral form and, for small frequencies and in the thermodynamic limit $L \to \infty$, we obtain a scaling regime as discussed below.
Furthermore, Eq. (\ref{eq:current_power_speck_scale}) shows that, as $\tilde{y}\to \infty$, the scaled power spectrum of instantaneous currents diverges with the system size as $2L-1$, which has been illustrated in Fig. \ref{fig:Power_speck_cur_MCM_I}.
The above mentioned scaling function for the power spectrum of instantaneous bond current as a function of scaled frequency $\tilde{y}$ has an integral representation in the lower frequency regime $D/L^2 \ll f \ll 1$. The integral representation is as follows:
\begin{align}
    \mathcal{H}(\tilde{y}) \simeq\lim\limits_{L\to \infty}\frac{L}{\pi}\int\limits_{2\pi/L}^{\pi} dq \left[1+\frac{\omega(q)}{2}\right]\frac{1}{1+\frac{L^4\omega(q)^2}{4\pi^2\tilde{y}^2}},
\end{align}
where $\omega(q)=2(1-\cos q)$. Now after variable transformation $z = \omega(q)L^2$ and then by taking $L\to\infty$,  we obtain the following expression of the scaling function,
\begin{equation}\label{eq:assy_sj}
    \mathcal{H}(\tilde{y}) \simeq\frac{1}{2\pi}\int\limits_{0}^{\infty}\frac{dz}{z^{1/2}[1+\frac{z^2}{4\pi^2\tilde{y}^2}]}\simeq\sqrt{\frac{\tilde{y}\pi}{4}}.
\end{equation}
In Fig. \ref{fig:Power_speck_cur_MCM_I}, one can see that the scaled power spectrum of the instantaneous bond current, denoted as $L\tilde{S}_{\cj}(f)/(2\chi)$, plotted against the scaled frequency $L^2f/D$ for various chipping parameters and system sizes, all at a global density of $\rho=1$. The simulation results are depicted by using solid colored lines, while the black solid line represents theoretical predictions from Eq.(\ref{eq:current_power_speck_scale}), which agree very well with the simulation data. Additionally, in the lower scaled-frequency range, we have included a guiding line representing $\tilde{y}^{1/2}$ behavior, as specified in Eq.(\ref{eq:assy_sj}).
According to Eq.(\ref{eq:assy_sj}), the power spectrum of instantaneous current displays a power-law behavior $f^{\psi_{\cj}}$ with an exponent $\psi_{\cj}=1/2$ in the low-frequency regime. Indeed, it can be immediately inferred that, in the temporal domain, the correlation of instantaneous current has a scaling behavior of $\langle \mathcal{J}_0(t)\mathcal{J}_0(0) \rangle \sim t^{-\psi_{\mathcal{J}} - 1}$, i.e., a $t^{-3/2}$ power law decay, which is already obtained in Eq. \eqref{power-law1}.

\subsection{Subsystem mass fluctuations}

In the previous sections, we have performed a detailed study of dynamic current-current correlation and the associated power spectrum for instantaneous current. In this section, we study the power spectrum for subsystem mass fluctuation. to this end, we first exactly calculate the two-point dynamic correlation function  $C_r^{mm}(t,0)=\abr{m_i(t)m_{i+r}(0)}-\abr{m_i(t)}\abr{m_{i+r}(0)}$ for local mass. By employing the previously mentioned microscopic update rules, we can derive the time evolution of $C_r^{mm}(t,0)\equiv C_r^{mm}(t)$,  
\begin{align}
\begin{aligned}\label{eq:cmm_1}
    \frac{d}{dt}C_r^{mm}(t)&= D\sum\limits_{k}\Delta_{0,k}\abr{m_k(t)m_{r}(0)}_c\\
    &=D\sum\limits_{k}\Delta_{0,k}C_k^{mm}(t).
    \end{aligned}
\end{align}
The solution of the above equation can be written in terms of Fourier representation, 
\begin{align}\label{eq:c_mm2}
    \tilde{C}_q^{mm}(t) = e^{-D\omega_qt}\tilde{C}_q^{mm}(0),
\end{align}
where $\tilde{C}_q^{mm}$ is the Fourier transform of $C_r^{mm}$. The equal-time two-point mass correlation function $C_r^{mm}(0)$ corresponds to the steady-state mass-mass correlation $C_r^{mm}$ mentioned in Eq.(\ref{eq:mass-mass_f}). Notably, the equal-time mass correlation $C_r^{mm}$ has a direct connection with the scaled subsystem-mass fluctuation,
\begin{align}\label{eq:sig_m}
    \sigma_M^2 \equiv \lim \limits_{l\to\infty}\frac{\abr{M_l^2}-\abr{M_l}^2}{l}=\sum\limits_{r=-\infty}^{r=\infty}C_r^{mm} =\frac{\chi}{D},
\end{align}
where boundary contribution of $C_r^{mm}$ has been neglected in the limit of large subsystem size $l\to\infty$.

Note that, in Eq. \eqref{eq:sig_m}, the density-dependent transport coefficient - the collective particle mobility $\chi(\rho)$ - is defined purely from current fluctuations in the systems, where the particle-hopping rates are strictly symmetric in either directions. Indeed, the essence of the recently developed macroscopic fluctuation theory for diffusive systems \cite{Bertini2015Jun} is that, for ``gradient-type models'', the current fluctuations can be alternatively calculated using a slightly different approach, where the particle-hopping rates are biased in a certain direction. Essentially, this amounts to applying a small biasing force in that direction so that the hopping rates become slightly asymmetric and a small current is generated in the system. Interestingly, this particular scheme leads to definition of another transport coefficient, which we call an ``operational'' mobility and characterizes the response (i.e., the small current generated) of the system to a small biasing force field $F$; for simplicity, the biasing field is assumed to be constant throughout. Indeed, in Ref. \cite{Arghya_2017}, one of us previously introduced such a biasing force, which modifies the original unbiased (symmetric) hopping rates of the MCMs. Of course, in the absence of the force $F$, we recover the original time-evolution equation (\ref{eq:generic_diff_eqn})). In that case, a time-evolution equation for local density field, as opposed to the unbiased scenario as in Eq.(\ref{eq:generic_diff_eqn})), is given by \cite{Arghya_2017}
\begin{align}
    \frac{d\rho_i}{dt} = \frac{\lamt}{2}(\rho_{i+1}-2\rho_i+\rho_{i-1}) +\frac{\lamt^2}{12}F\left(\abr{m_{i-1}^2}-\abr{m_{i+1}^2}\right),
\end{align}
where we denote local density $\rho_i(t)=\abr{m_i(t)}$ at site $i$. By scaling space and time as $x=i/L$ and $\tau=t/L^2$ and the biasing force as $F =\tilde{F}/L$, the time-evolution equation for density field $\rho(x,\tau)$ as a function of the rescaled space and time  variables can be expressed in terms of a continuity equation,
\begin{align}\label{eq:hydro}
\begin{aligned}
    \partial_{\tau}\rho(x,\tau) &= -\partial_x\left[-D\partial_x\rho(x,\tau)+\chi_{op} \tilde{F}\rho(x,\tau)\right]\\
    &=-\partial_x J(x,\tau),
    \end{aligned}
\end{align}
where the total local current $J=J_{diff}+J_{drift}$ is written as the sum of the diffusive current $J_{diff}=-D\partial_x\rho(x,\tau)$ and drift current $J_{drift}=\chi_{op}\tilde{F}$; here the two transport coefficients - the bulk-diffusion coefficient and the ``operational" mobility are given by $D=\lamt/2$ and  $\chi_{op}=\abr{m_i^2}/6$, respectively. The latter identity immediately implies, directly through Eq.(\ref{eq:mobility:mcm_1}), a fluctuation-response  relation between the operational mobility and the current fluctuation,
\begin{equation}
\chi_{op}(\rho) \equiv \left[\frac{\partial J_{drift}}{\partial \tilde{F}}\right]_{\tilde{F}=0} = \sigma_Q^2 \equiv \chi (\rho).
\end{equation} 
In other words, here we have derived a {\it nonequilibrium} version of the celebrated Green-Kubo relation for (near) equilibrium systems. We can immediately derive a nonequilibrium version of another celebrated relation for systems in equilibrium, called the Einstein relation, which connects the scaled mass fluctuation, the bulk-diffusion coefficient and the ``operational'' mobility, i.e.,
\begin{align} \label{ER}
    \chi_{op}\equiv\left[\frac{\partial J_{drift}}{\partial \tilde{F}}\right]_{\tilde{F}=0} = D\sigma^2_M,
\end{align}
where we have used the already derived fluctuation relation as given in Eq.(\ref{eq:sig_m}). Notably, the above equation is exact for the MCMs studied in this paper and the above analysis constitutes a microscopic derivation of the fluctuation-response relation. Furthermore, by using Eq.(\ref{eq:sig_Q}) and Eq. (\ref{eq:sig_m}), we can immediately derive another  nonequilibrium fluctuation relation, between fluctuation of mass and that of current, as expressed in the following equation,
\begin{align}\label{eq:mass_cur_fluc}
\sigma^2_{M}=\frac{\sigma^2_{\mathcal{Q}}}{2D}.
\end{align}
It is not difficult to see that the above relation is nothing but a slightly modified version of the equilibrium-like Einstein relation as given in Eq. \eqref{ER}. While the above set of fluctuation relations are quite well established in the context of equilibrium systems, their existence in systems having a nonequilibrium steady state is however nontrivial and is not rigorously proven for many-particle systems having a nonequilibrium steady state. Indeed, a general theoretical understanding has been gradually emerging for a somewhat restricted class of nonequilibrium systems having diffusive bulk dynamics, which violate detailed balance in the bulk and these MCMs belong to. Indeed it is quite desirable to prove such fluctuation relations through exact microscopic approaches, which are still a formidable task even  for these simplest class of many-particle model systems studied here.

In order to solve Eq.(\ref{eq:c_mm2}), we must determine the steady-state mass-mass correlation in Fourier modes, denoted as $C_q^{mm}$ given below,
\begin{align}
    C_q^{mm}  = \frac{\chi}{D}\left(1+\frac{\omega_q}{2}\right).
\end{align}
Now, we substitute the above  into Eq.(\ref{eq:c_mm2}) to obtain the solution,
\begin{align}
    \tilde{C}_q^{mm}(t) = \frac{\chi}{D}e^{-D\omega_qt}\left(1+\frac{\omega_q}{2}\right).
\end{align}
Finally, by using inverse Fourier transformation, we get the correlation function in real time,
\begin{align}\label{eq:crmm_1}
    C_r^{mm}(t) = \frac{\chi}{D}\frac{1}{L}\sum\limits_q e^{-iqr}e^{-D\omega_qt}\left(1+\frac{\omega_q}{2} \right).
\end{align}
To calculate the asymptotic behavior for large time, we express the above expression by setting $r=0$ and writing it in an integral form as
\begin{align}
    C_0^{mm}(t) \simeq \frac{\chi}{\pi D}\int\limits_0^{\pi} dq e^{-D\omega(q)t}\left[1+\frac{\omega(q)}{2}\right].
\end{align}
Now, we approximate $\omega(q)\approx q^2$, perform a variable transformation $z = D t q^2$ and simplify the above equation as
\begin{align}
     C_0^{mm}(t) \simeq \frac{\chi}{\pi D\sqrt{4Dt}}\int\limits_{0}^{\pi}z^{-\frac{1}{2}}e^{-z}dz,
\end{align}
where the subleading term $O(t^{-3/2})$ is neglected. After putting the value of the integral $\int_{0}^{\infty}z^{-\frac{1}{2}}e^{-z}dz = \sqrt{\pi}$ in the above equation, we obtain the desired asymptotic expression,
\begin{align}
    C_0^{mm}(t) \simeq \frac{\chi}{D\sqrt{4\pi D}}t^{-\frac{1}{2}}.
\end{align}
We now consider a subsystem of size $l<L$ with a total mass $M_l(t)=\sum_{i=0}^{l-1}m_i(t)$ and calculate the unequal-time correlation function $C^{M_lM_l}(t,0)\equiv C^{M_lM_l}(t)$ for subsystem mass as following,
\begin{align}
    C^{M_lM_l}(t) = \abr{\sum\limits_{i=0}^{l-1}m_i(t)\sum\limits_{j=0}^{l-1}m_j(0)}_c.
\end{align}
Upon simplifying the above equation, we obtain the following identity
\begin{align}
    C^{M_lM_l}(t) = lC_0^{mm}(t) + \sum\limits_{r=1}^{l-1}(l-r)\left[C_r^{mm}(t)+C_{-r}^{mm}(t)\right].
\end{align}
After substituting Eq.(\ref{eq:crmm_1}) into the above equation and by performing some algebraic manipulations, we arrive at the expression given below:
\begin{align}\label{eq:c_mlml_r}
    C^{M_lM_l}(t) = \frac{\chi}{D}\frac{1}{L}\sum\limits_q e^{-D\omega_qt}\left(1+\frac{\omega_q}{2} \right)\frac{\omega_{lq}}{\omega_q}.
\end{align}
Now, we derive the asymptotic behavior of the dynamic correlation function $C^{M_lM_l}(t)$ for subsystem mass, which appeared in Eq.(\ref{eq:c_mlml_r}). At time $t=0$, it takes a maximum value and, subsequently, it decays as a function of time $t$. To extract the time dependence, we isolate $C^{M_lM_l}(t)$ by subtracting it from its maximum value and then express the equation in an approximate integral form as
\begin{align}
\begin{aligned}
  &C^{M_lM_l}(0) - C^{M_lM_l}(t)
  \\
  & \simeq \frac{2\chi}{\pi D}\int\limits_{0}^{\pi}dq \left[1-e^{-D\omega(q)t}\right]\left[1+\frac{\omega(q)}{2}\right]\frac{1}{\omega (q)}.
    \end{aligned}
\end{align}
Again, by approximating $\omega(q)\approx q^2$ and performing a variable transformation $z = D t q^2$, we simplify the above equation in the leading order as
\begin{align}
    C^{M_lM_l}(0) - C^{M_lM_l}(t)\simeq \frac{\chi \sqrt{t}}{\pi \sqrt{D}}\int\limits_0^\infty z^{-\frac{3}{2}}(1-e^{-z})dz.
\end{align}
By using $\int_{0}^{\infty}z^{-3/2}(1-e^{-z})dz = 2\sqrt{\pi}$, we obtain the asymptotic expression of dynamic correlation function for the subsystem mass,
\begin{align}
    C^{M_lM_l}(t) - C^{M_lM_l}(0)\simeq -\frac{2\chi}{\sqrt{\pi D}}t^{\frac{1}{2}}.
\end{align}
Now we can also calculate the power spectrum of subsystem mass fluctuation $S_{M_l}(f)$ by calculating the Fourier transform of the Eq.(\ref{eq:c_mlml_r}),
\begin{equation}
\begin{aligned}\label{eq:smf_mcm_I_1}
    S_{M_l}(f)&=\lim\limits_{T\to\infty}\int\limits_{-T}^{T}dt   ~C^{M_lM_l}(t,0)e^{2\pi ift}.
    \end{aligned}
\end{equation}
Upon completing the aforementioned integration, we derive the desired expression for the power spectrum,
\begin{align}\label{eq:mass_fluc_MCM_I_f}
    S_{M_l}(f) =\frac{2\chi}{L}\sum\limits_{q}\left(1+\frac{\omega_q}{2}\right)\frac{\omega_{lq}}{\omega_q^2D^2+4\pi^2f^2},
\end{align}
where $\omega_q=2(1-\cos q)$, with $q=2\pi n/L$ and $n=1,2,\cdots, L-1$.
We then obtain another scaling function, denoted as $\mathcal{F}$, that connects power spectrum $S_{M_l}(f)$ to a scaled frequency in the following manner:
\begin{equation}\label{eq:mass_fluc_scale}
\begin{aligned}
    \mathcal{F}\left(\frac{L^2f}{D}\right)&=\frac{D^2}{2\chi L^3}S_{M_l}(f)\\
    &=\lim_{L\to\infty}\sum\limits_{q}\left(1+\frac{\omega_q}{2}\right)\frac{\omega_{lq}}{L^4\omega_q^2+4\pi^2(L^2f/D)^2}.
    \end{aligned}
\end{equation}
It is worth noting that the subsystem-mass power spectrum $S_{M_l}(f)$ can be well approximated through the following integral,
\begin{align}
    \mathcal{F}(\tilde{y}) &\simeq\lim\limits_{L\to \infty}\frac{L}{\pi}\int\limits_{2\pi/L}^{\pi} dq \left[1+\frac{\omega(q)}{2}\right]\frac{\omega(lq)}{4\pi^2\tilde{y}^2+L^4\omega(q)^2}.
\end{align}
For large subsystem size $1 \ll l < L$, the function $\omega(lq)$ exhibits high-frequency oscillations with values in the range of $[0,4]$, leading to an approximation of $\omega(lq) \approx 2$. Additionally, we can make use of the transformation $z = \omega(q)L^2$ and then by taking $L\to\infty$, we obtain the following expression of the scaling function,
\begin{align}
    \mathcal{F}(\tilde{y}) \simeq\frac{1}{4\pi^3\tilde{y}^2}\int\limits_{0}^{\infty}\frac{dz}{z^{1/2}\left(1+\frac{z^2}{4\pi^2\tilde{y}^2}\right)}.
\end{align}
After performing the above integral, we obtain the asymptotic expression of the scaled power spectrum of the subsystem mass fluctuation,
\begin{equation}\label{eq:sM_l_asym}
    \mathcal{F}(\tilde{y}) \simeq \frac{(\tilde{y}\pi)^{-\frac{3}{2}}}{4}.
\end{equation}
The above equation implies that, in low-frequency regime, the power spectrum of the subsystem mass displays a power-law behavior $f^{-\psi_M}$ with an exponent of $\psi_M = 3/2$. It is also evident that, in the temporal domain, the correlation of the subsystem mass $\langle M_l(t)M_l(0) \rangle_c$ scales as $t^{\psi_M-1}$, which is, interestingly, a $t^{1/2}$ power-law growth.

\begin{figure}[H]
        \centering
        \includegraphics[scale=0.5]{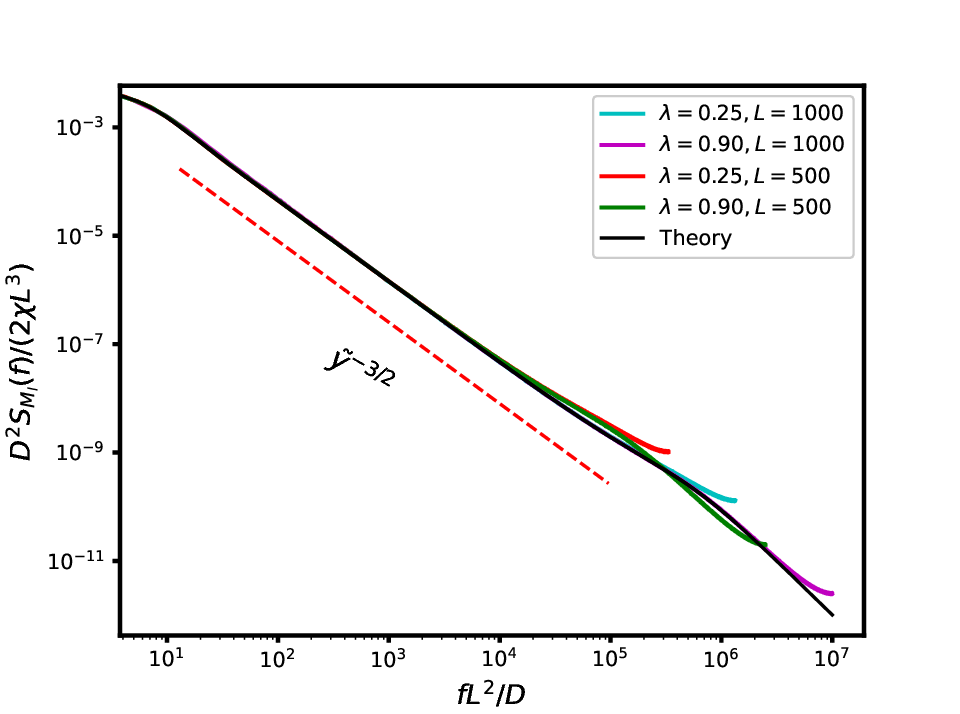}
        \caption{The scaled subsystem mass power spectrum, $D^2\tilde{S}_{M_l}(f)/(2\chi L^3)$, is plotted as a function of the scaled frequency $L^2f/D$ for various chipping parametre and system sizes. The lines, colored cyan ($\lambda=0.25$, $L=1000$), magenta ($\lambda=0.90$, $L=1000$), red ($\lambda=0.25$, $L=500$), and green ($\lambda=0.90$, $L=500$), represent simulation data, all obtained at a global density of $\rho=1$.
        In the low-frequency range, the red dashed line demonstrates a scaling behavior of $\mathcal{F}(\tilde{y})$ as $\tilde{y}^{-3/2}$ [as in Eq.(\ref{eq:sM_l_asym})]. The black solid line corresponds to theoretical predictions [as mentioned in Eq.(\ref{eq:mass_fluc_scale})], providing an excellent match with the simulation data for that system size.}
        \label{fig:msf_msm_I_diff_lam}
    \end{figure}

In Fig. \ref{fig:msf_msm_I_diff_lam}, we plot the scaled power spectrum $D^2\tilde{S}_{M_l}(f)/(2\chi L^3)$ for subsystem mass as a function of the scaled frequency $L^2f/D$ for various chipping parameters and system sizes, along with theoretical predictions provided there as black solid line for comparison purpose. The power spectrum exhibits a power-law scaling in the low-frequency range, with the red dashed line indicating a $\tilde{y}^{-3/2}$ behavior. The simulation results (solid colored lines) are in excellent agreement with the theoretical predictions (black solid line).

    \section{Model: MCM II}
    \label{Sec:Model_II}
    
  In this section, we apply the theoretical framework developed in the previous sections. In the case of  MCM II, a site is selected randomly and a fraction of the chipped-off mass is transferred either to the right nearest neighbor or the left nearest neighbor, with the remaining chipped-off mass being deposited back onto the same site. This dynamic behavior results in the absence of nearest-neighbor correlations, thus simplifying the calculation of dynamic correlations. We now present the significant findings for the MCM II model by discussing the main results.

\subsection{Time-integrated current fluctuation}

The time-integrated bond-current fluctuation for MCM II is described by the expression as given in eq. (\ref{eq:QQ_t_0_mcmII}). In this case, the strength of the fluctuating current is characterized by $\Gamma_r=2\chi \delta_{0,r}$, where the bulk diffusion coefficient is given by $D=\tilde{\lambda}/4$ and the density-dependent mobility $\chi(\rho)=\lamt^2\rho^2/2(3-2\lamt)$ remains identical to that for MCM I [see Eqs. (\ref{eq:mobility:mcm_1}) and (\ref{eq:mi2})]. It is worth mentioning that, for MCM II, both the steady-state density correlation  $C_r^{mm}$ and the strength of the fluctuating current $\Gamma_r$ are delta-correlated, indicating the absence of nearest-neighbor correlations (see Table I for comparison between models); also $\Gamma_r$ and $C_r^{mm}$ are related by a scaling factor (the bulk diffusivity $D$) as given in Eq.(\ref{eq:gamm_r_cr}), and this particular relationship holds true for this model as well.

\begin{figure}[H]
    \centering
    \includegraphics[scale=0.5]{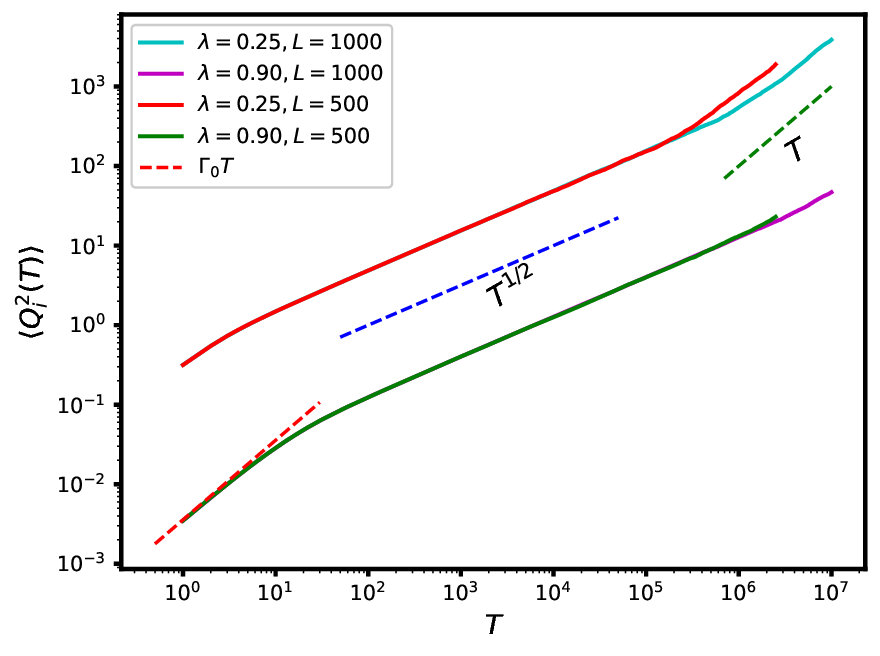}
    \includegraphics[scale=0.5]{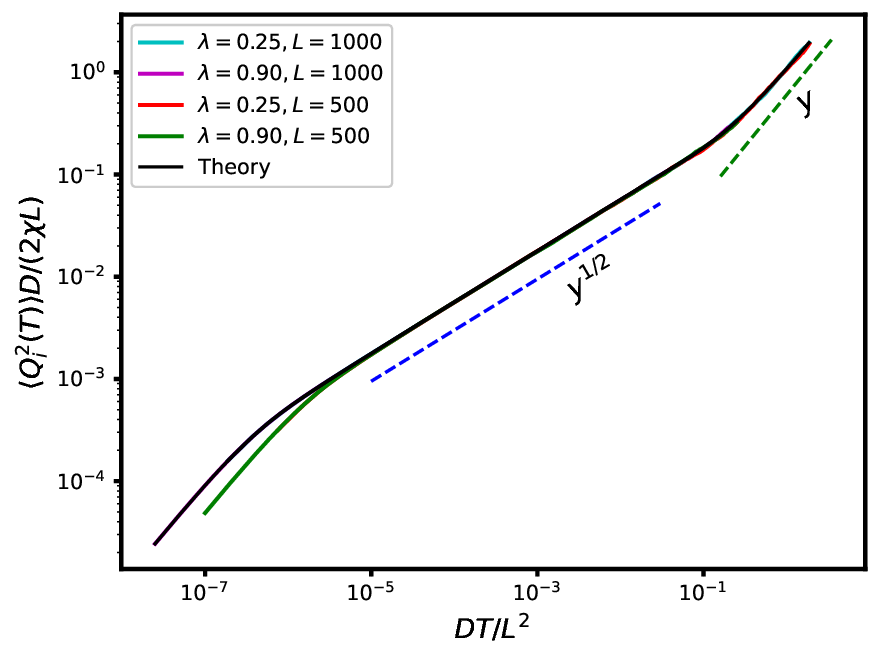}
    \caption{\textit{In top panel:}We present the time-integrated bond-current fluctuation, denoted as $\langle \mathcal{Q}_i^2(T) \rangle$, plotted as a function of time $T$ for various chipping parameters and system sizes. For both this panel, the cyan ($\lambda=0.25$, $L=1000$), magenta ($\lambda=0.90$, $L=1000$), red ($\lambda=0.25$, $L=500$), and green ($\lambda=0.90$, $L=500$) lines represent simulation data at a global density of $\rho=1$. The red dashed line corresponds to the behavior $\Gamma_0T$ for $\lambda = 0.90$ (refer to Eq.(\ref{eq:gamma_0t_II})), while the blue and green dashed lines represent sub-diffusive growth, approximately scaling as $\sim T^{1/2}$, and diffusive growth, approximately scaling as $\sim T$, respectively, as discussed in Eq.(\ref{eq:qi2t_tt_scalling_full_mcmII})
    \textit{In bottom panel:} The plot displays the scaled time-integrated bond-current fluctuation, $\langle \mathcal{Q}_i^2(T) \rangle D/(2\chi L)$, as a function of scaled time $DT/L^2$ for various chipping parameter and system sizes. The two dashed lines serve as guides, indicating that $\mathcal{W}(y)$ exhibits sub-diffusive behavior as $\sim y^{1/2}$ (in blue) at early times, followed by diffusive growth as $\sim y$ (in green) at later times [as in Eq.(\ref{eq:qi2t_tt_scalling})]. The solid color lines illustrate the simulation results, while the black solid line represents the theoretical results obtained from Eq.(\ref{eq:QQ_t_0_mcmII}) upon suitable scaling, and this theoretical curve perfectly matches the simulation data.}
    \label{fig:qi2t_diff_lam_mcm_II}
\end{figure}

For MCM II, we proceed to calculate the time-integrated bond-current fluctuation $C_0^{QQ}(t,t)$ for equal time $t' =t=T$ and in the same space, i.e., $r=0$. The resulting expression is as follows:
\begin{equation}\label{eq:QQ_t_0_mcmII}
    \langle \mathcal{Q}_i^2(T)\rangle = \frac{2\chi T}{L} +\frac{2\chi}{L}\sum\limits_{n=1}^{L-1}\frac{(1-e^{-D\omega_nT})}{D\omega_n} ,
\end{equation}
where $\omega_n=2(1-\cos(2\pi n/L))$, with $n=0,1,\cdots,L-1$. If we take the $T\to\infty$ limit first in the above equation, we get the following expression,
\begin{align}
     \langle \mathcal{Q}_i^2(T)\rangle \simeq \frac{2\chi T}{L}+ \frac{\chi L}{6D} = \frac{2\chi T}{L} \left[ 1 + {\cal O}\left( \frac{L^2}{DT} \right) \right].
\end{align}
It is worth noting that the term $\chi L/6D$ in the above equation can be neglected since the leading contribution will arise from the term $2\chi T/L$ as $T \gg L^2$.
In the smaller time regime where $DT \ll 1$, Eq.(\ref{eq:QQ_t_0_mcmII}) can be simplified as 
\begin{equation}\label{eq:gamma_0t_II}
     \langle \mathcal{Q}_i^2(T)\rangle=\Gamma_0 T = 2\chi T,
\end{equation}
where the strength of the fluctuating current $\Gamma_0$ is equal to $2\chi$, which differs from the value in the case of MCM I (as demonstrated in Eq.(\ref{eq:gamma_r_f})).
Furthermore, in the regime where $DT\gg 1$, we find that the scaling function $\mathcal{W}(y)$, with the scaling variable $y=DT/L^2$, has the exactly same form as presented in Eq.(\ref{eq:Scalling_cur}), as found in MCM I.
In top panel of Figure \ref{fig:qi2t_diff_lam_mcm_II}, we present the time-integrated bond-current fluctuation, denoted as $\langle \mathcal{Q}_i^2(T) \rangle$, as a function of time $T$. We highlight the early-time behavior of $ \langle \mathcal{Q}^2(T)\rangle$, which scales as $\Gamma_0T$ and is described in Eq.(\ref{eq:gamma_0t_II}). Additionally, in the bottom panel of the figure, we plot the scaled fluctuation $\langle \mathcal{Q}_i^2(T) \rangle D/(2\chi L)$ of the time-integrated bond current as a function of the scaled time $DT/L^2$. We include guiding lines that illustrate the asymptotic behavior of the scaling function $\mathcal{W}(y)$.
Thus the overall behavior for the temporal growth of the time-integrated bond-current fluctuation exhibits three distinct (asymptotic) regimes:
\begin{equation}\label{eq:qi2t_tt_scalling_full_mcmII}
\langle\mathcal{Q}_i^2(T)\rangle=
\begin{cases}
   2\chi T & \text{for}~ DT\ll  1\\
   \frac{2\chi}{\sqrt{D\pi}}T^{\frac{1}{2}} & \text{for}~ 1 \ll DT \ll L^2 \\
   \frac{2\chi T}{L} & \text{for} ~DT\gg L^2.
\end{cases}
\end{equation}

\subsection{Instantaneous bond-current fluctuations}

\label{subsec:Powspeck_cur_MCM_II}

Using the theory presented in Section \ref{Sec:Theory}, we also compute the power spectrum of the instantaneous bond current $\mathcal{J}_i(t)$ in MCM II. The expression for the power spectrum can be written as
\begin{equation}\label{eq:current_power_speck_MCM_II}
    \tilde{S}_\cj(f)=\frac{2\chi(\rho)}{L}\sum\limits_{q}\frac{4\pi^2f^2}{D^2\omega_q^2+4\pi^2f^2},
    \end{equation}
where, $\omega_q=2(1-\cos q)$, with $q=2\pi n/L$ and $n=1,2,\cdots, L-1$.
The above expression exhibits an asymptotic behavior $\mathcal{H}(\tilde{y})\sim \tilde{y}^{1/2}$ in the low-frequency regime, quite similar to that for MCM I as mentioned in Eq.(\ref{eq:assy_sj}). Furthermore, $\mathcal{H}(\tilde{y})$ diverges as $L-1$ in the limit $\tilde{y}\to \infty$ (i.e., for $T \gg L^2$).

\begin{figure}[H]
    \centering
    \includegraphics[scale=0.5]{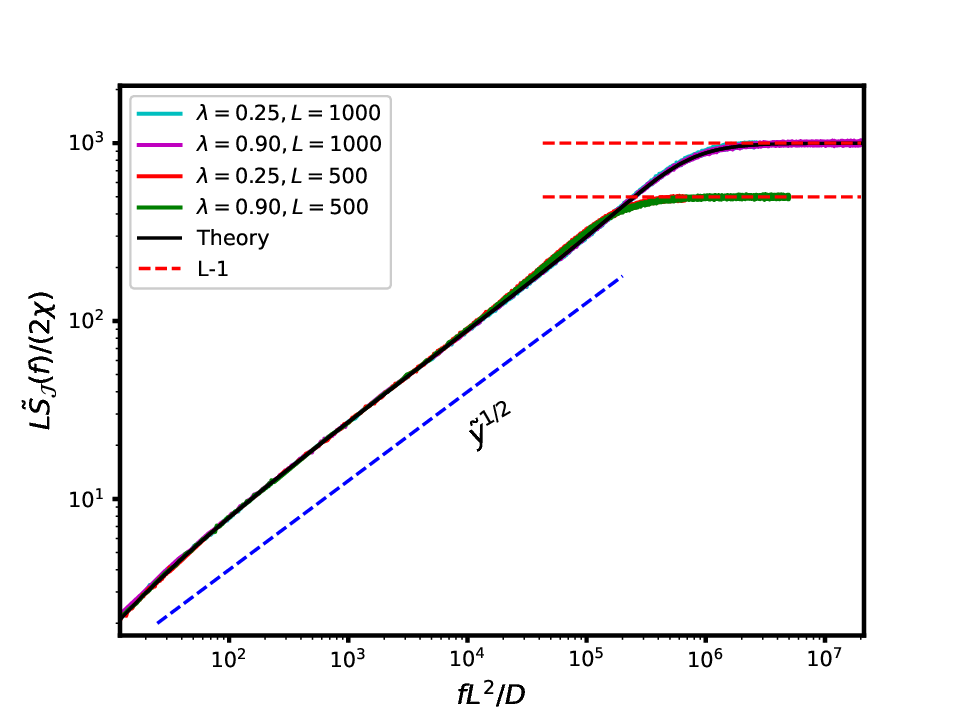}
    \caption{The scaled power spectrum of instantaneous
    currents, $L\tilde{S}_{\cj}(f)/(2\chi)$, is plotted as a function
    of scaled frequency $L^2f/D$ for various chipping parameters and system sizes.
    The cyan solid line corresponds to $\lambda=0.25$ and $L=1000$, the magenta solid line corresponds to $\lambda=0.90$ and $L=1000$, the red solid line corresponds to $\lambda=0.25$ and $L=500$, and the green solid line corresponds to $\lambda=0.90$ and $L=500$, all at a global density of $\rho=1$.
    The blue dashed line shows $\tilde{y}^{1/2}$ scaling behavior in the low-frequency regime as in Eq.(\ref{eq:assy_sj}) and red dashed lines represent $L\tilde{S}_{\cj}(f)/(2\chi)$ diverges as system size $L-1$ at the high-frequency limit. The solid color lines represent the simulation results, while the black solid line represents the theoretical predictions of Eq.(\ref{eq:current_power_speck_MCM_II}) upon suitable scaling, which fully matches the simulation data.}
    \label{fig:Power_speck_cur_MCM_II}
\end{figure}

In Figure \ref{fig:Power_speck_cur_MCM_II}, the scaled power spectrum $L\tilde{S}_{\cj}(f)/(2\chi)$ of the instantaneous bond current is plotted against the scaled frequency $L^2f/D$. Simulation and theoretical results are found to have an excellent agreement with each other. Additionally, in the low-frequency range, we have included a guiding line $\sim \tilde{y}^{1/2}$, which is obtained from the integral representation of Eq.(\ref{eq:current_power_speck_MCM_II}). This behavior is consistent with what was mentioned earlier in the context of MCM I.

\subsection{Subsystem mass fluctuations}

\begin{figure}[H]
    \centering
    \includegraphics[scale=0.5]{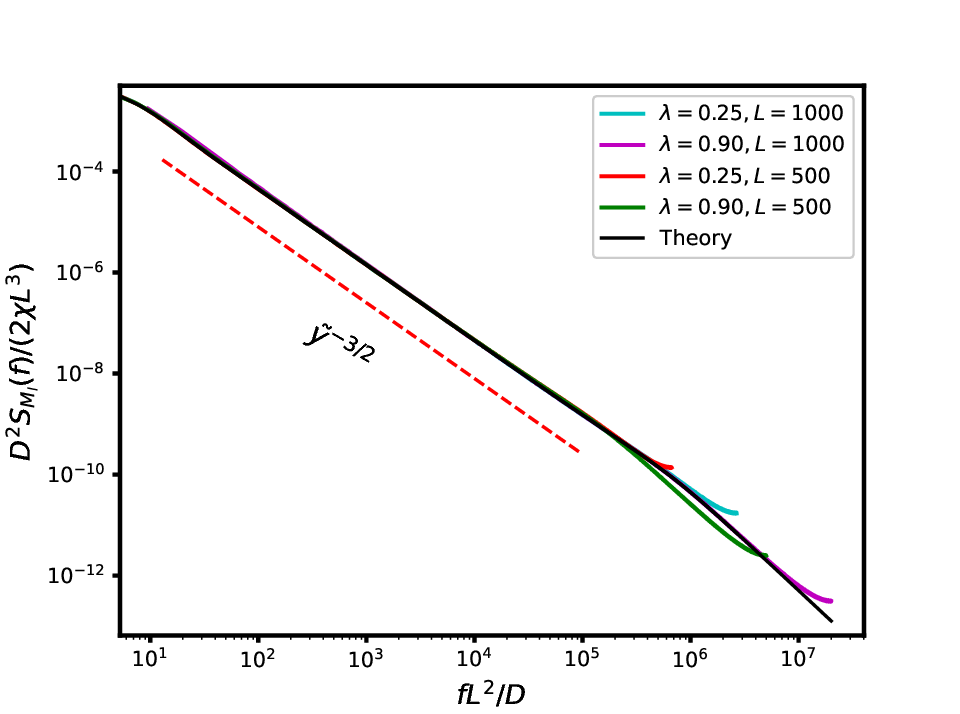}
    \caption{The scaled subsystem mass power spectrum, $D^2\tilde{S}_{M_l}(f)/(2\chi L^3)$, is plotted as a function of scaled frequency $L^2f/D$ for various chipping parametre and system sizes. The cyan solid line corresponds to $\lambda=0.25$ and $L=1000$, the magenta solid line corresponds to $\lambda=0.90$ and $L=1000$, the red solid line corresponds to $\lambda=0.25$ and $L=500$, and the green solid line corresponds to $\lambda=0.90$ and $L=500$, all at a global density of $\rho=1$. The red dashed line shows a $\tilde{y}^{-3/2}$ scaling behavior in the low-frequency regime. The solid color lines represent the simulation results, while the black solid line represents the theoretical predictions of Eq.(\ref{eq:mass_fluc_MCM_II}) upon suitable scaling, which matches the simulation data.}
    \label{fig:msf_msm_II_diff_lam}
\end{figure}

For MCM II, we compute another time-dependent quantity - the  power spectrum $S_{M_l}(f)$ for subsystem mass,
\begin{equation}\label{eq:mass_fluc_MCM_II}
    S_{M_l}(f) =\frac{2\chi}{L}\sum\limits_{q}\frac{\omega_{vq}}{\omega_q^2D^2+4\pi^2f^2}.
    \end{equation}
We can now calculate the scaling function $\mathcal{F}(\tilde{y})$ associated with the above mentioned power spectrum of subsystem mass. Notably, it exhibits a $\tilde{y}^{-3/2}$ power-law behavior in the low-frequency regime, similar to that observed in the case of MCM I mentioned in Eq.(\ref{eq:sM_l_asym}).
In Figure \ref{fig:msf_msm_II_diff_lam}, the scaled subsystem-mass power spectrum $D^2\tilde{S}_{M_l}(f)/(2\chi L^3)$ is plotted against the scaled frequency $L^2f/D$ for various chipping parameters and system sizes. The red dashed line exhibits a $\tilde{y}^{-3/2}$ power-law behavior in the low-frequency regime. The solid color lines represent the simulation results and the black solid line corresponds to the theoretical predictions from Eq.(\ref{eq:mass_fluc_MCM_II}), which matches quite well with the simulation data when suitably scaled.

\section{MCM III}\label{Sec:MCM_III}

In this section, we compute the dynamic correlations for current and mass in the case of MCM III, which has been extensively investigated in the past, to explain wealth distribution in a population \cite{Vledouts2015Dec,Chakraborti2000Sep}. In this model, each site retains a fraction of its mass and the remaining mass is distributed among its nearest neighbor sites. The distribution of the mixed masses among these sites is randomized. It is important to note that, upon appropriate scaling, the dynamic correlations for MCM III turn out to be similar to those for model MCM II, arising from the fact that both these models exhibit similar density correlations (with vanishing neighboring correlations) due to their specific microscopic dynamics.

\subsection{Time-integrated current fluctuation}

The time-integrated bond current fluctuation in MCM III takes an identical form to that of MCM II, as shown in Eq.(\ref{eq:QQ_t_0_mcmII}). Indeed the strength  $\Gamma_0$  of the fluctuating current and the particle $\chi$ are the same as in MCM II. Also the bulk diffusivity $D=\frac{\tilde{\lambda}}{2}$ for MCM III is again independent of density.

\begin{figure}[H]
    \centering
    \includegraphics[scale=0.5]{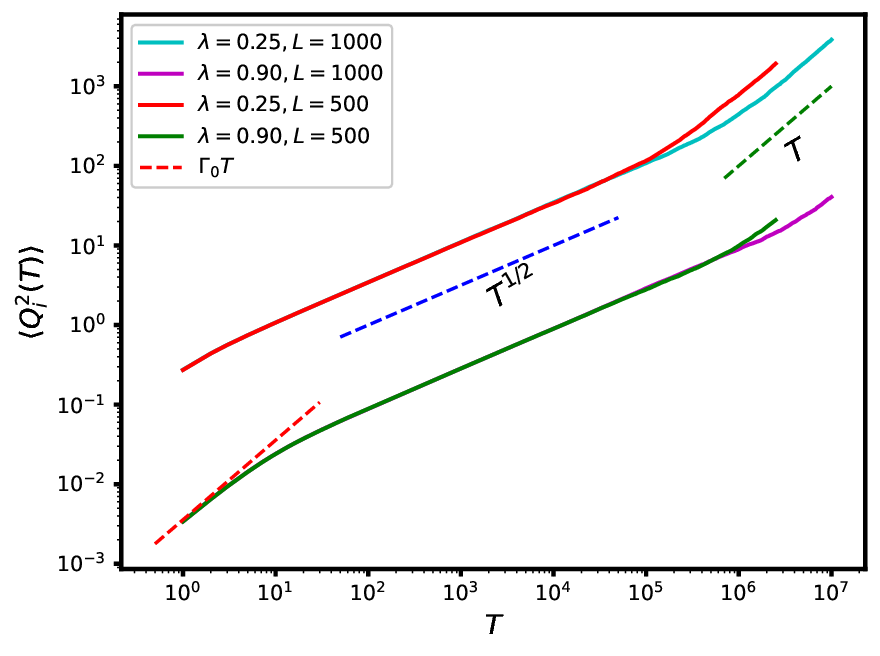}
    \includegraphics[scale=0.5]{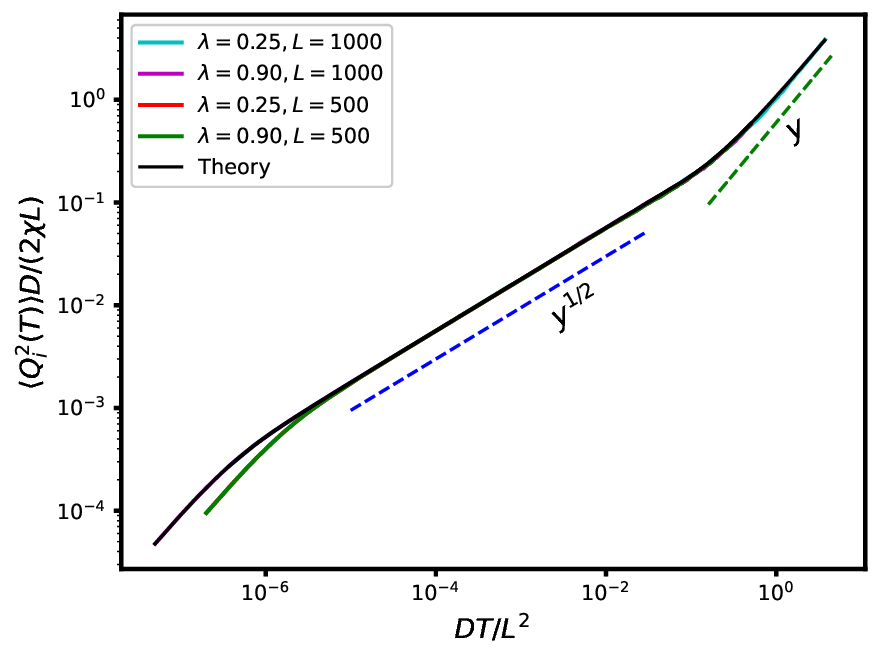}
    \caption{\textit{In top panel:} The time-integrated bond-current fluctuation, $\langle \mathcal{Q}_i^2(T) \rangle$, plotted as a function of time $T$ for various chipping parameter and system sizes. For both this panel, the cyan ($\lambda=0.25$, $L=1000$), magenta ($\lambda=0.90$, $L=1000$), red ($\lambda=0.25$, $L=500$), and green ($\lambda=0.90$, $L=500$) lines lines represent simulation data for global density $\rho=1$. The red dashed line represents the behavior $\Gamma_0T$ for $\lambda = 0.90$ [see Eq(\ref{eq:gamma_0t_II})], while the blue and green dashed lines represent sub-diffusive $\sim T^{1/2}$ and diffusive $\sim T$ growth, respectively, as mentioned in Eq.(\ref{eq:qi2t_tt_scalling_full_mcmII}).
    \textit{In bottom panel:} The plot shows the scaled time-integrated bond-current fluctuation, $\langle \mathcal{Q}_i^2(T) \rangle D/(2\chi L)$, as a function of scaled time $DT/L^2$ for different chipping parameter and system sizes. The two dashed lines serve as guides, indicating that $\mathcal{W}(y)$ exhibits sub-diffusive behavior as $\sim y^{1/2}$ (in blue) at early times, followed by diffusive growth as $\sim y$ (in green) at later times [as in Eq.(\ref{eq:qi2t_tt_scalling})]. Solid color lines represent simulation results, while the black solid line represents the theoretical prediction obtained from Eq.(\ref{eq:QQ_t_0_mcmII}), which perfectly matches the simulation data when suitably scaled.}
    \label{fig:qi2t_MCM_III}
\end{figure}

In  top panel of Figure \ref{fig:qi2t_MCM_III}, we present the time-integrated bond-current fluctuation, denoted as $\langle \mathcal{Q}_i^2(T) \rangle$, as a function of time $T$. The early-time behavior of $\langle \mathcal{Q}_i^2(T)\rangle$, which scales as $\Gamma_0T$ according to Eq.(\ref{eq:gamma_0t_II}), is highlighted. In the bottom panel, we display the scaled fluctuation of the time-integrated current, $\langle \mathcal{Q}_i^2(T) \rangle D/(2\chi L)$, as a function of scaled time $DT/L^2$. Solid color lines represent simulation results, while the black solid line represents the theoretical prediction obtained from Eq.(\ref{eq:QQ_t_0_mcmII}), which agrees well with the simulation data when appropriately scaled. Guiding lines illustrate the asymptotic behavior of the scaling function $\mathcal{W}(y)$.

\subsection{Instantaneous bond-current fluctuations}

\label{subsec:Powspeck_cur_MCM_III}

\begin{figure}[H]
    \centering
    \includegraphics[scale=0.5]{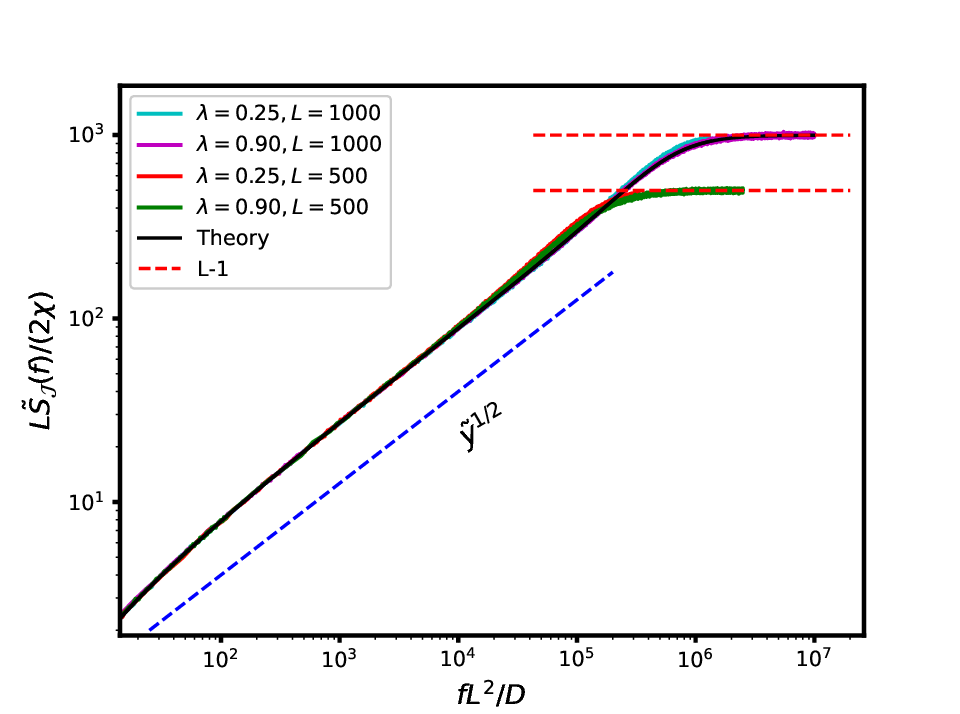}
    \caption{The scaled power spectrum of instantaneous
    currents, $L\tilde{S}_{\cj}(f)/(2\chi)$, is plotted as a function
    of scaled frequency $L^2f/D$ for various chipping parameters and system sizes.
    The cyan solid line corresponds to $\lambda=0.25$ and $L=1000$, the magenta solid line corresponds to $\lambda=0.90$ and $L=1000$, the red solid line corresponds to $\lambda=0.25$ and $L=500$, and the green solid line corresponds to $\lambda=0.90$ and $L=500$, all at a global density of $\rho=1$.
    The blue dashed line shows $\tilde{y}^{1/2}$ scaling behavior in the low-frequency regime and the red dashed lines represent the power spectrum diverges as $L-1$ at the high-frequency limit. The solid color lines represent the simulation results, while the black solid line represents the theoretical predictions of Eq.(\ref{eq:current_power_speck_MCM_II}) upon suitable scaling, which fully matches the simulation data.}
    \label{fig:Power_speck_cur_MCM_III}
    \end{figure}

For MCM III, the power spectrum of instantaneous bond current exhibits the same form as mentioned in Eq.(\ref{eq:current_power_speck_MCM_II}), with $D=\lambda/2$ and the particle mobility $\chi(\rho)=\lamt^2\rho^2/2(3-2\lamt)$ being the same as above mentioned MCMs (MCM I and MCM II). 
In Fig. \ref{fig:Power_speck_cur_MCM_III}, the scaled power spectrum  $L\tilde{S}_{\cj}(f)/(2\chi)$ of the instantaneous bond current is plotted against the scaled frequency $L^2f/D$. Both simulation and theoretical results are shown to have an excellent agreement with each other. Additionally, in the low-frequency range, we have included a guiding line $\sim \tilde{y}^{1/2}$; this particular behavior is consistent with what was mentioned earlier in the context of MCM I [see Eq.(\ref{eq:assy_sj})].

\subsection{Subsystem Mass fluctuations}

\begin{figure}[H]
    \centering
    \includegraphics[scale=0.5]{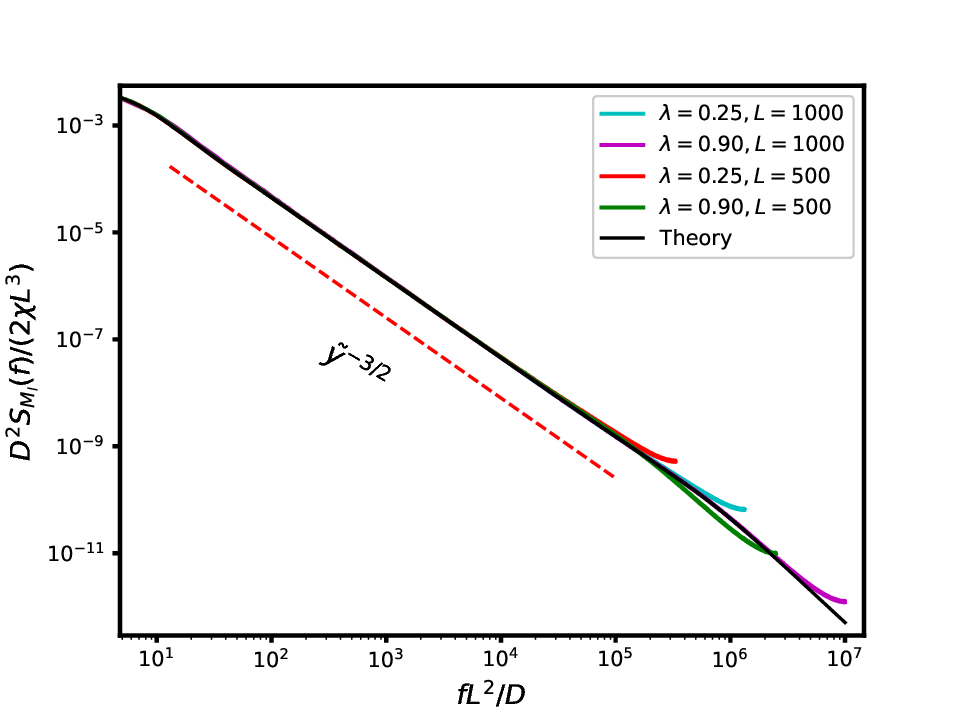}
    \caption{The scaled subsystem mass power spectrum, $D^2\tilde{S}_{M_l}(f)/(2\chi L^3)$, is plotted as a function of scaled frequency $L^2f/D$ for various chipping parameter and system sizes. The cyan solid line corresponds to $\lambda=0.25$ and $L=1000$, the magenta solid line corresponds to $\lambda=0.90$ and $L=1000$, the red solid line corresponds to $\lambda=0.25$ and $L=500$, and the green solid line corresponds to $\lambda=0.90$ and $L=500$, all at a global density of $\rho=1$. The red dashed line shows a $\tilde{y}^{-3/2}$ scaling behavior in the low-frequency regime. The solid color lines represent the simulation results, while the black solid line represents the theoretical predictions of Eq.(\ref{eq:mass_fluc_MCM_II}) upon suitable scaling, which matches the simulation data.}
    \label{fig:msf_msm_III_diff_lam}
\end{figure}

In this section, we discuss the subsystem-mass power spectrum, which has the same form as in Eq.(\ref{eq:mass_fluc_MCM_II}). In Figure \ref{fig:msf_msm_III_diff_lam}, we plot the scaled subsystem mass power spectrum, denoted as $D^2\tilde{S}_{M_l}(f)/(2\chi L^3)$, against the scaled frequency $L^2f/D$ for various chipping parameters and system sizes. Solid color lines represent simulation results, and the black solid line corresponds to theoretical predictions from Eq.(\ref{eq:mass_fluc_MCM_II}), in agreement with the simulation data when suitably scaled. The red dashed line shows a $\tilde{y}^{-3/2}$ scaling behavior at low frequencies, as in MCM I [see Eq.(\ref{eq:sM_l_asym})].

\section{Comparison of models}

\label{Sec:Comparison}

\begin{figure*}[t]
  \centering
  {\includegraphics[width=1\textwidth]{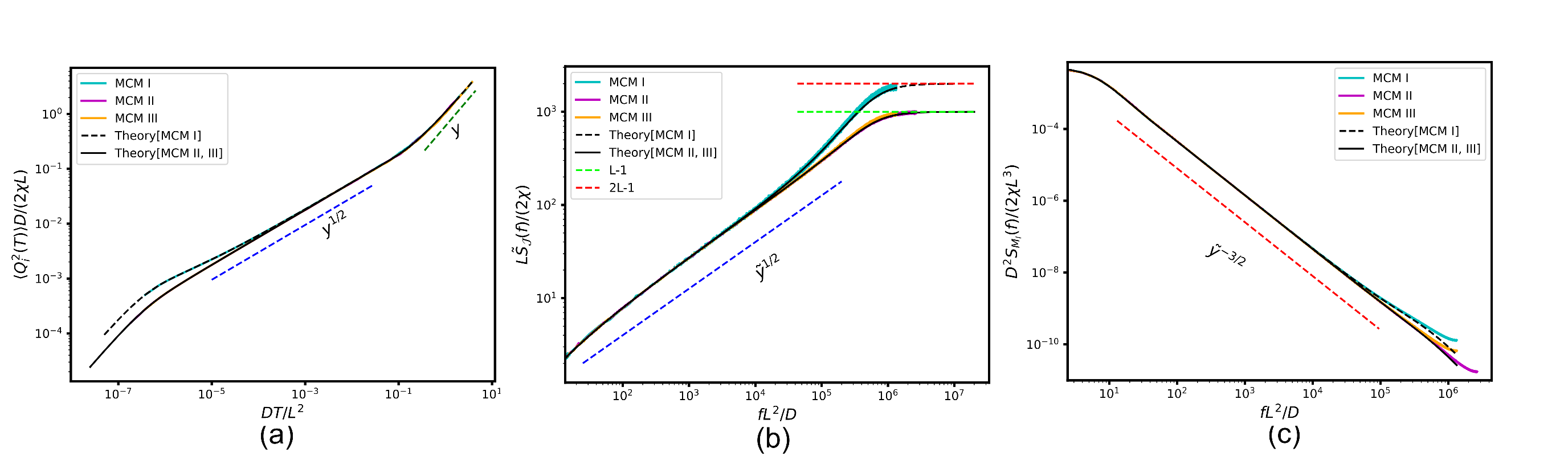}}
  \caption{(a) Scaled time-integrated bond-current fluctuation, $\langle \mathcal{Q}_i^2(T) \rangle D/(2\chi L)$, as a function of scaled time $DT/L^2$ for three models: MCM I (blue solid line), MCM II (magenta solid line), and MCM III (orange solid line). The two dashed lines serve as guides, indicating that $\mathcal{W}(y)$ exhibits sub-diffusive behavior as $\sim y^{1/2}$ (in blue) at early times, followed by diffusive growth as $\sim y$ (in green) at later times. (b) Scaled power spectrum of instantaneous currents, $L\tilde{S}_{\mathcal{J}}(f)/(2\chi)$, plotted as a function of scaled frequency $L^2f/D$ for various chipping parameter and system sizes. The blue dashed line represents the $\tilde{y}^{1/2}$ scaling behavior of $\mathcal{H}(\tilde{y})$ in the low-frequency regime. In contrast, the power spectrum exhibits divergence, reaching $2L-1$ (red dashed line) for MCM I and $L-1$ (lime dashed line) for MCM II and MCM III in the high-frequency limit. (c) Scaled subsystem mass power spectrum, $D^2\tilde{S}_{M_l}(f)/(2\chi L^3)$, plotted as a function of scaled frequency $L^2f/D$ for various chipping parameters and system sizes. The red dashed line shows a $\tilde{y}^{-3/2}$ scaling behavior of $\mathcal{F}(\tilde{y})$ in the low-frequency regime. In all of these panels, we have used a fixed chipping parameter $\lambda=0.5$, global density $\rho=1.0$, and a system size of $L=1000$. Simulation data for MCM I, MCM II, and MCM III are represented by blue, magenta, and orange solid lines, respectively. Black dashed lines correspond to MCM I theory, while black solid lines represent MCM II and MCM III theory, which are identical.}
     \label{fig:compare_Models}
    \end{figure*}

In this section, we perform a comparative study of various dynamical quantities for the three models studied in this paper: MCM I, MCM II, and MCM III. In particular, we do comparative investigation of the scaled time-integrated bond-current fluctuation $\mathcal{W}(y)$ as a function of scaled time $y=DT/L^2$, the scaled power spectrum of instantaneous currents $\mathcal{H}(\tilde{y})$, and the scaled subsystem mass power spectrum $\mathcal{F}(\tilde{y})$ as a function of scaled frequency $\tilde{y}=fL^2/D$.
In Fig. (\ref{fig:compare_Models}), to compare the model systems, we present  plots of these dynamical quantities as a function of the scaled time and frequencies. Remarkably, despite the different microscopic dynamical rules in each of the systems, we find that there exists a universal scaling regime, where all three models in fact exhibit the same behavior and the subdiffusive and diffusive grwoths of bond current fluctuations are connected through a single and the same scaling function. This particular observation suggests certain universal features, which are shared among these models, and presumably diffusive systems in general, even when the underlying microscopic dynamical rules are completely different.
However, outside the scaling regime, deviations are observed among the models. These deviations are accurately captured in both the simulation data and theoretical predictions, providing insights into the distinct dynamical properties of each system at small times.

    \begin{table*}
    \label{table:compare}
        {\renewcommand{\arraystretch}{2}%
    \begin{tabular}{| m{3.8cm} | m{5.9cm} | m{3.4cm} | m{3.4cm} | }
       \hline
       \textbf{Quantity} & \textbf{MCM I} & \textbf{MCM II} & \textbf{MCM III} \\
       \hline
       Bulk-diffusivity: \textbf{D} & $ \frac{\tilde{\lambda}}{2}$ 
       & $ \frac{\tilde{\lambda}}{4}$  
       &   $\frac{\tilde{\lambda}}{2}$ \\
       \hline
       Mobility: \textbf{$\mathbf{\chi}$} & $ \frac{\tilde{\lambda}^2\rho^2 }
       {2(3-2\tilde{\lambda} )}$ & 
       $\frac{\tilde{\lambda}^2\rho^2 }{2(3-2\tilde{\lambda} )}$ 
       & $\frac{\tilde{\lambda}^2\rho^2 }{2(3-2\tilde{\lambda} )}$ \\
       \hline
       Density correlation: $\mathbf{C_r^{mm}}$ & $\frac{\tilde{\lambda }\rho^2 }{2(3-2\tilde{\lambda})}[4\delta_{0,r}-\delta_{r,1}-\delta_{r,-1}]$ &
       $\frac{2\tilde{\lambda}}{(3-2\tilde{\lambda})}\rho^2\delta_{0,r}$ 
       & $\frac{\tilde{\lambda}}{(3-2\tilde{\lambda})}\rho^2\delta_{0,r}$ \\
       \hline
       $\mathbf{\Gamma_r}$ & $\frac{\tilde{\lambda }^2\rho^2 }{2(3-2\tilde{\lambda})}[4\delta_{0,r}-\delta_{r,1}-\delta_{r,-1}]$ 
       & $\frac{\tilde{\lambda}^2\rho^2 }{(3-2\tilde{\lambda} )}\delta_{r, 0}$
        & $\frac{\tilde{\lambda}^2\rho^2 }{(3-2\tilde{\lambda} )}\delta_{r,0}$\\
       \hline
       $\mathbf{A_r}$ & $\frac{\tilde{\lambda }^2\rho^2 }{4(3-2\tilde{\lambda})}[-5(\delta_{r,0}-\delta_{r,-1})+(\delta_{r,1}-\delta_{r,-2})]$ & $-\frac{1}{2}\frac{\lamt^2\rho^2}{(3-2\lamt)}
        (\delta_{r,0}-\delta_{r,-1})$ & $-\frac{1}{2}\frac{\lamt^2\rho^2}{(3-2\lamt)}
        (\delta_{r,r}-\delta_{r,-1})$ \\
       \hline
       $\mathbf{\tilde{f}_q}$ & $ -\frac{\tilde{\lambda}^2\rho^2 }
       {2(3-2\tilde{\lambda} )}(1-e^{-iq})\Big(1+\frac{w_q}{2} \Big)$ & $-\frac{\tilde{\lambda}^2\rho^2 }
       {2(3-2\tilde{\lambda} )}(1-e^{-iq})$ & $-\frac{\tilde{\lambda}^2\rho^2 }
       {2(3-2\tilde{\lambda} )}(1-e^{-iq})$ \\
        \hline
       
       \end{tabular}}\quad
   
    \caption{We have highlighted key quantities related to dynamical correlations in these three models: MCM I, MCM II, and MCM III. These quantities include transport coefficients such as bulk diffusivity $D$ and mobility $\chi$. Additionally, we have mentioned steady-state density correlation $C_r^{mm}$, the strength of fluctuating current $\Gamma_r$, the source term in the time evolution equation of mass-current correlation at equal times $A_r$, and its Fourier mode $\tilde{f}_q$. These quantities are essential for deriving dynamical correlations in these models.}
   
 \end{table*}

We provide in Table I a concise description concerning the similarities and differences among the three models, in terms of their time-dependent properties. To begin with, we highlight the transport coefficients - the bulk-diffusion coefficient $D$ and the particle mobility $\chi(\rho)$ for each model. Notably, $D$ is constant (independent of density) for all three models; however, the mobility $\chi$ is density dependent (proportional to square of the density). The table displays the density correlation function $C_r^{mm}$ for these models. MCM I possesses finite (nonzero) nearest-neighbor spatial correlations, whereas MCM II and MCM III lack such correlations. The table also presents the strength $\Gamma_r$ of fluctuating (``noise'') current for the models, with each of the models having a distinct value. However, it is important to note that the relationship $\Gamma_r=C_r^{mm}/2D$ holds true for all models; presumably this relation is valid for diffusive systems in general. Lastly, the table compares the source term $A_r$ appearing in the time-evolution equation for mass-current correlation function, where the Fourier mode $\tilde{f}_q$ of the quantity $A_r$ is useful in explicitly calculating various dynamic quantities).

\section{summary and Conclusion}
\label{Sec:Conclusion}

In this paper, we exactly calculate dynamic correlation functions for mass and current in a broad class of conserved-mass transport processes, called mass chipping models (MCMs) on a one dimensional ring. These systems - variants of intensively studied  random average processes (RAPs) \cite{Ferrari1998}  - violate detailed balance in the bulk and, unlike symmetric simple exclusion processes (SSEP) and zero range processes (ZRP), have nontrivial spatial structures even on a periodic lattice. Indeed, in most cases  [a notable exception being the Kipnis-Marchioro-Presutti (KMP) model \cite{Kipnis1982Jan}],  their steady-state measures (on a ring) are not described by the equilibrium Boltzmann-Gibbs distribution and a priori {\it not} known. 
Qualitatively we find three temporal growth regimes for the fluctuation of the time-integrated bond current. For initial times and for all three models (MCMs I, II and III), the time-integrated current fluctuation grows linearly with time $T$, with a proportionality factor of $\Gamma_0$, which, however, is a model-dependent quantity and determined exactly for each of the models. In the intermediate, but large, time regime $1/D \ll T \ll L^2/D$, with $D$ being the bulk diffusion coefficient, we find subdiffusive - $T^{1/2}$ - growth of the bond-current fluctuation, where the density-dependent prefactor of $T^{1/2}$ growth are calculated. This subdiffusive grwoth is again followed by a linear, or diffusive, growth of current fluctuation in the long-time regime (when $T \gg L^2$). Furthermore, using a microscopic approach, we exactly calculate a {\it model-independent} scaling function $\mathcal{W}(y) \equiv D \langle \mathcal{Q}_i^2(T) \rangle / 2\chi L$ as a function of a scaling variable $y = DT/L^2$, where $D$, $\chi$ and $L$ are the bulk-diffusion coefficient, mobility, and system size $L$, respectively. It is quite interesting that a single scaling function indeed connects both the intermediate-time subdiffusive and long-time diffusive growths of the time-integrated bond-current fluctuation.

 We also analytically calculate the dynamic correlation function for instantaneous bond current. We show that, similar to the SSEP  \cite{Sadhu_2016}, the correlation function for bond current decays as $t^{-3/2}$ even if the MCMs, unlike the SSEP  on a periodic domain, have  nontrivial spatial structures in the bulk. Notably, the correlation function has a delta-correlated part at $t=0$ and its magnitude for $t>0$ is in fact negative Despite the fact that there is no restriction on single-site occupancy of mass, i.e., mass at a site is unbounded (unlike the SSEP), the negative part of the current correlation function is directly responsible for the subdiffusive growth of the bond current fluctuation in the thermodynamic limit. The power-law behavior of dynamic current correlations is consistent with the exact calculation of the current power spectrum $\tilde{S}_{\mathcal{J}}(f)$, which has a low-frequency asymptotic behavior $\tilde{S}_{\mathcal{J}}(f) \sim f^{\psi_\mathcal{J}}$ with $\psi_\mathcal{J} = 1/2$. Furthermore, we exactly obtain the scaling function which $\mathcal{H}(\tilde{y})$, which represents the rescaled power spectrum of the current $L \tilde{S}_{\mathcal{J}}(f) / 2\chi$  as a function of the scaled variable $\tilde{y} = fL^2 / D$.

We also calculate the scaled subsystem mass fluctuation, which is shown to be identically equal to the suitably scaled dynamic fluctuation of the space-time integrated current $\mathcal{Q}_{sub}(l,T)$, divided by factor of $2D$, with $D$ being the bulk-diffusion coefficient [see Eq.(\ref{eq:mass_cur_fluc})]. This particular fluctuation relation in the context of MCMs is nothing but a nonequilibrium version of the celebrated equilibrium Einstein relation. 
It should be noted that the fluctuation relation requires the intensive fluctuation of space-time integrated current to be calculated in the thermodynamic limit, i.e., by first taking the limit of the infinite subsystem size limit $l \rightarrow \infty$, followed by the limit of infinite time $T \rightarrow \infty$. In this specific order of limits, we also show that  $\lim_{l \rightarrow \infty} \lim_{T \rightarrow \infty} \langle \mathcal{Q}_{sub}^2(l,T) \rangle/ lT$ is identically equal to the spatial sum of the strength $\Gamma_r$ of the fluctuating bond current. As a simple consequence of the Einstein relation, the correlation functions of mass and fluctuating current are related by $C_r^{mm} = \Gamma_r / 2D$. We also calculate the unequal-time correlation function of the total mass of a subsystem and  its power spectrum $S_{M_l}$, which decays as $S_{M_l} \sim f^{-\psi_M}$, with $\psi_M = 3/2$, consistent with the scaling relation $\psi_\mathcal{J} = 2 - \psi_M$ \cite{Anirban_2023}. The scaled power spectrum of subsystem mass, $D^2 S_{M_l}(f) / 2\chi L^3$, can also be expressed in terms of a scaling function $\mathcal{F}(\tilde{y})$ with scaling variable $\tilde{y} = fL^2 / D$.

Notably, the qualitative behavior of time-integrated bond-current fluctuations for all three models are similar, though the prefactors of the temporal growth laws depend on the details of the dynamical rules. In the small-frequency (large-time) domain, the prefactors of the intermediate-time subdiffusive and long-time diffusive growth of the bond current fluctuations can be expressed in terms of the bulk-diffusion coefficient and the particle mobility. However, the large-frequency behavior is not universal in the sense that they depend on the local microscopic properties. More specifically, the differences in the spatial structures of these  models manifest themselves in the small-time growth of the time-integrated bond-current fluctuation $\langle \mathcal{Q}^2_i(T) \rangle \simeq \Gamma_0 T$, where $\Gamma_0$ is proportional to the single-site mass fluctuation $\langle m_i^2 \rangle$, which is different in these three models.

A few remarks are in order. Characterizing the dynamic properties of interacting-particle systems through microscopic calculations is an important, though difficult, problem in statistical physics.  The variants of mass chipping models discussed here have nontrivial steady states, but they are shown to be still analytically tractable and exact results concerning dynamic correlations could be obtained. As discussed in this paper, depending on their dynamical rules, these models differ from each other in the details of their spatial structures. Model MCM I possesses nonzero spatial correlations, whereas MCM II and III have vanishing neighboring correlations. However all these models share one noteworthy aspect in common: The bulk-diffusion coefficient, like in the SSEP  \cite{Sadhu_2016}, is {\it independent} of density. In fact, this is precisely why the hierarchy of current and mass correlations closes and thus the models are exactly solvable. Although the mass chipping models have a nontrivial steady state,  these models, due to their simple dynamical rules,  do not exhibit a phase transition, or there are no singularities present in the transport coefficients. However, through simple variations in the dynamical rules, it is possible to have nontrivial (singular) macroscopic behavior in some of their variants. Indeed, it would be quite interesting to characterize the dynamic properties of current and mass through microscopic calculations in higher dimensions, for systems with more than one conserved quantity \cite{Basile2006May} and when  the transport properties are singular or anomalous \cite{Anirban_2023}.

\section*{Acknowledgement}

We thank Tanmoy Chakraborty, Arghya Das and Anupam Kundu for helpful discussions at various stages of the project. P.P. gratefully acknowledges the Science and Engineering Research Board (SERB), India, under Grant No. MTR/2019/000386, for financial support. A.M. acknowledges financial support from the Department of Science and Technology, India [Fellowship No. DST/INSPIRE Fellowship/2017/IF170275] for part of the work carried out under his senior research fellowship.

\bibliography{ref.bib}

\end{document}